%% file: main.tex
\documentclass[letterpaper]{article}
\usepackage{aaai2027}
\nocopyright
\usepackage[hyphens]{url}
\usepackage{graphicx}
\usepackage{natbib}
\usepackage{caption}
\usepackage{booktabs}
\usepackage{tikz}
\usepackage{array}

\title{TCA-SIR: Learning Target-Conditioned Abstractions for Scientific Inspiration Retrieval}
\author{
Yuto Suzuki\textsuperscript{\rm 1}\thanks{Corresponding author.},
Farnoush Banaei-Kashani\textsuperscript{\rm 1}
}
\affiliations{
\textsuperscript{\rm 1}Department of Computer Science and Engineering,
University of Colorado Denver\\
Denver, Colorado, USA\\
\{yuto.suzuki,farnoush.banaei-kashani\}@ucdenver.edu
}

\begin{document}

\maketitle

\input{sections/abstract}

\input{sections/introduction}
\input{sections/related_work}

\input{sections/problem_definition}
\input{sections/method}
\input{sections/experiments}
\input{sections/results}
\input{sections/conclusion}

\bibliography{refs}

\clearpage
\appendix
\input{appendix/overview}
\input{appendix/additional_results}
\input{appendix/implementation_details}
\input{appendix/teacher_supervision_quality}
\input{appendix/prompts}
\input{appendix/training_examples}

\end{document}

%% file: sections/abstract.tex
\begin{abstract}
Scientific hypothesis generation is a key component of the emerging AI for Science systems. Hypothesis generation is performed in two steps: 1) \emph{Scientific Inspiration Retrieval (SIR)}, to identify the existing inspiring knowledge that can seed a hypothesis to address a target problem, and 2) \textit{Hypothesis Composition}, to compose the hypothesis for addressing the target problem based on the retrieved inspirations. Existing SIR methods rank papers merely based on topical similarity without explicitly representing how a candidate inspiration is transferable to the target. This approach particularly disregards remote inspirations (i.e., inspirations from knowledge domains other than the target domain), which are known for generation of most creative hypotheses. The relevance of such inspirations depends not on topical similarity but on reusable problem-solving principles they offer. Motivated by how humans transfer ideas by abstracting transferable aspects of source inspirations and remapping them to address a target problem, we reformulate SIR as \emph{target-conditioned abstraction} (TCA). With this approach, the retrieved object is a transferable abstract principle extracted from a candidate inspiration specifically relevant for addressing the target problem. Accordingly, we present \textbf{TCA-SIR}, a novel TCA-based SIR method that learns to generate target-conditioned abstractions for candidate inspirations and uses their learned representations to predict the transferability of the inspirations to the target. Using the ResearchBench benchmark, we demonstrate that TCA-SIR outperforms the existing SIR solutions as well as direct LLM retrieval, improving HitRate@top4\% over the state-of-the-art MOOSE-Chem method by more than 10 percentage points. Comparisons with the same backbone using an untrained TCA prompt further show that learned abstractions more clearly recover target-relevant mechanisms (including chemical levers and performance bottlenecks) rather than topical slogans or superficial facts. By explicitly exposing the reusable principles connecting a candidate inspiration to the target, TCA-SIR not only improves retrieval performance, but also provides an \textit{interpretable} rationale for scientific inspiration retrieval.
\end{abstract}

%% file: sections/introduction.tex
\section{Introduction}

AI for science is advancing rapidly, driven by both scientific opportunity and capability gains in large language models (LLMs).
Systems such as the AI Scientist series automate substantial parts of the discovery pipeline and can draft papers with limited human intervention~\citep{lu2024ai,yamada2025ai,lu2026towards}.
AI already contributes at each stage of the scientific discovery process~\citep{wei2025ai}: (1)~observation and hypothesis generation~\citep{si2025can,romera2024mathematical,yang2026moose,novikov2025alphaevolve,liu2026towards,radensky2026scideator,baek2025researchagent}, (2)~experimental planning and execution~\citep{boiko2023autonomous,swanson2024virtual}, (3)~result analysis~\citep{wang2025spatialagent,ding2024automating}, and (4)~synthesis, validation, and evolution~\citep{ou2025claimcheck,takagi2023towards}.
Among these stages, hypothesis generation is especially consequential: a weak hypothesis wastes downstream experimental and analytical effort, yet robust, scalable hypothesis generation remains an open problem.

\begin{figure*}[t]
\centering
\includegraphics[width=\textwidth]{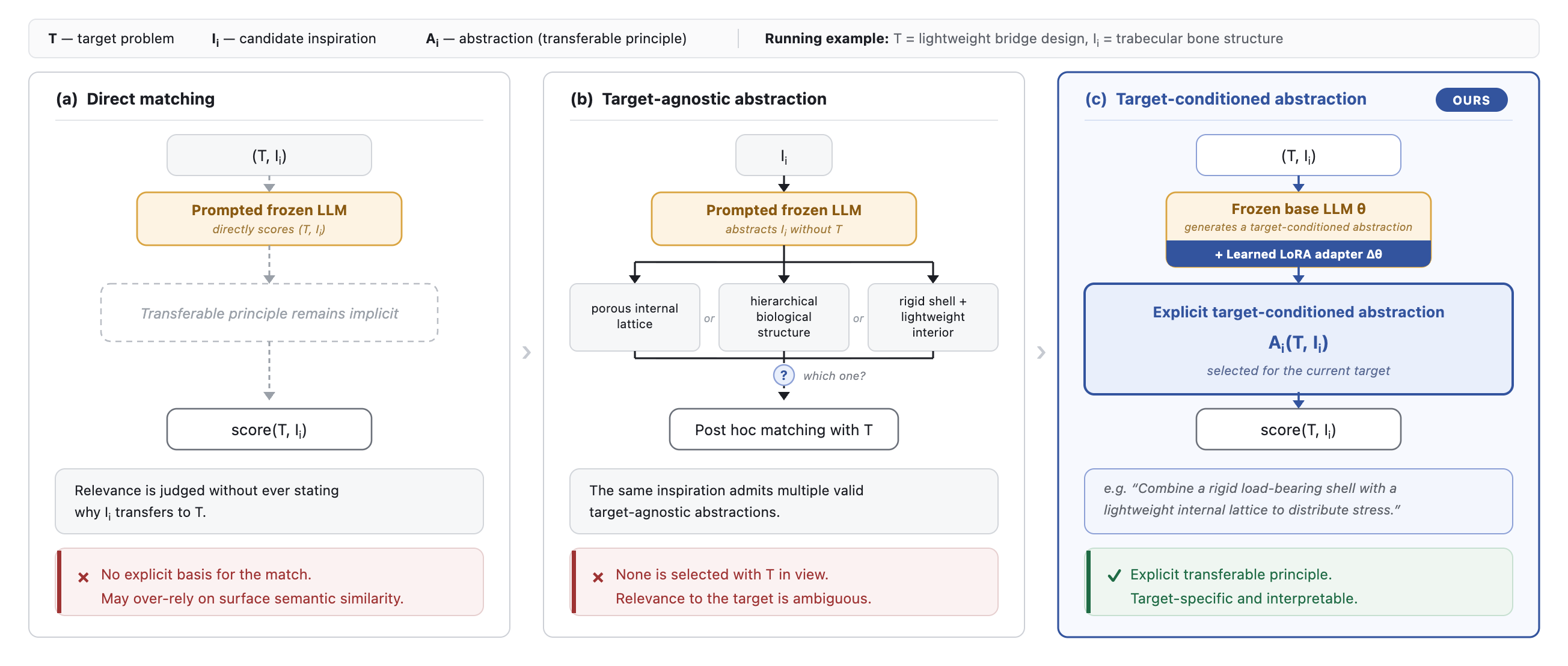}
\caption{Three formulations of scientific inspiration retrieval.
(a)~direct matching leaves the transferable principle implicit~\citep{yang2025moose};
(b)~target-agnostic abstraction produces multiple possible cores~\citep{gu2024llms,kargupta2026sparking};
(c)~target-conditioned abstraction selects the principle relevant to the target, making retrieval explicit and interpretable.}
\label{fig:concept-comparison}
\end{figure*}

MOOSE-Chem~\citep{yang2025moose} and ResearchBench~\citep{liu2026researchbench} further decompose hypothesis generation into \emph{scientific inspiration retrieval} (SIR) and hypothesis composition.
SIR asks systems to rank papers that can seed a hypothesis for a target research problem.
This step is particularly difficult to implement for LLMs because useful inspirations often come from distant domains and share little topic similarity with the target problem. Such inspirations must be recognized by identifying transferable abstract principles/ideas rather than topical overlap.
Prior work attacks this difficulty by enriching embeddings with citation-graph context~\citep{garikaparthi2025mir}, generating multi-level abstractions to strip domain-specific detail~\citep{gu2024llms}, or extracting domain-agnostic core problems and matching papers that share them~\citep{kargupta2026sparking}.
These approaches help, but they still leave a central ambiguity unresolved: a single inspiration admits many transferable readings along both \emph{facet} (e.g., a component, an analytical method, or an entire framework) and \emph{depth} (domain-specific vs.\ domain-agnostic statements).
Source-side abstraction alone is therefore insufficient, and useful retrieval must select the abstraction that connects to the \emph{target}.
Figure~\ref{fig:concept-comparison} contrasts this target-conditioned view with direct paper matching and target-agnostic abstraction.

Drawing on how humans apply remote analogies (by abstracting a portable core and remapping it to a new setting), we redefine SIR as \emph{target-conditioned abstraction} (TCA). With TCA, the retrieval object is not the paper itself, but a transferable principle extracted from the candidate and conditioned on the target.
We propose \textbf{TCA-SIR}, which jointly learns to generate Reasoning/Abstraction intermediates and to score transferability, so that the model learns which target-conditioned reading of an inspiration is useful.
Using ResearchBench~\citep{liu2026researchbench} as benchmark, rigorous comparisons against direct LLM retrieval and multiple abstraction-based baselines, together with ablations and qualitative analyses, show that TCA-SIR improves HitRate and ranking metrics while yielding interpretable mechanisms for why an inspiration helps a target, a useful signal for the subsequent composition stage.

We summarize our contributions as follows:
\begin{itemize}
    \item We redefine scientific inspiration retrieval as TCA (\emph{target-conditioned abstraction}), where the retrieval object is a transferable principle selected for the target rather than the candidate paper itself.
    \item We propose \textbf{TCA-SIR}, which jointly learns target-conditioned Reasoning/Abstraction generation and transferability scoring for ranking candidate inspirations.
    \item Through quantitative evaluation, ablations, and qualitative analysis with an established benchmark, we show that learned TCA improves retrieval over strong LLM baselines and provides interpretable explanations of inspirations' usefulness.
\end{itemize}

%% file: sections/related_work.tex
\section{Related Work}

\subsection{Scientific Inspiration Retrieval}
Computational approaches to scientific discovery have evolved from analogy mining~\citep{hope2017accelerating} to recent LLM-based scientific discovery systems such as SciMON~\citep{wang2024scimon} and CHIMERA~\citep{sternlicht2026chimera}.
MOOSE-Chem~\citep{yang2025moose} and ResearchBench~\citep{liu2026researchbench} further decompose hypothesis generation into \emph{inspiration retrieval} (ranking papers that can seed a hypothesis for a target problem) and hypothesis composition.
This decomposition highlights that the quality of the retrieved inspirations strongly influences downstream hypothesis generation: many useful inspirations are not background knowledge for the target domain, yet can still unlock a solution when recombined with what the researcher already knows.
That view aligns with classic accounts of creativity as unfamiliar combinations of familiar ideas~\citep{koestler1964act,boden2004creative}, and it helps explain why surface lexical overlap and domain-local retrieval often miss the inspirations that matter.
Existing retrieval methods address this challenge only partially.
MIR~\citep{garikaparthi2025mir} proposes related Methodology Inspiration Retrieval, which requires citation contexts and citation-graph signals to enrich embeddings. However, these artifacts are often unavailable or incomplete in many open candidate pools.
Combinatorial-creativity approaches~\citep{gu2024llms} generate multi-level abstractions (e.g., L1--L4) to enable cross-domain matching, yet the abstraction levels are heuristically defined, and matching often reduces to embedding or LLM similarity that need not reflect problem--solution transfer.
Idea-Catalyst~\citep{kargupta2026sparking} instead strips domain surface form to recover a domain-agnostic conceptual challenge and then seeks external papers that share that challenge; however, the critical extraction step relies entirely on prompting rather than task-specific learning.
Across these lines of work, inspiration retrieval benefits from some form of abstraction or restructuring, but prior methods leave the transferable core underspecified, unconditioned on the target, or dependent on side information that ResearchBench-style SIR settings do not provide.
Our work addresses this gap by learning target-conditioned transferable abstractions optimized directly for scientific inspiration retrieval.

\subsection{Abstraction Learning}
A growing recent line of work treats abstraction as part of the LLM reasoning process.
Abstraction-of-Thought~\citep{hong2024abstraction} inserts abstract thinking steps so models can reason at a higher level before specializing to a concrete solution.
RLAD~\citep{qu2025rlad} trains models to discover abstractions first and then follow more efficient reasoning paths.
AbstRaL~\citep{gao2025abstral} reinforces abstract thinking with the same goal of improving reasoning performance.
These methods primarily learn \emph{procedural} abstractions, i.e., compressed reasoning schemas that help solve the instance at hand.
They are not designed to extract a reusable conceptual core from one problem (or inspiration) so that it can be transferred to a different target. However, this is exactly the form of abstraction needed for scientific inspiration retrieval, and deriving such abstractions is one of the objectives of our work.

%% file: sections/problem_definition.tex
\section{Problem Definition}

\paragraph{Scientific Inspiration Retrieval in previous work.}
Prior work (namely, MOOSE-Chem~\citep{yang2025moose} and ResearchBench~\citep{liu2026researchbench}) defines Scientific Inspiration Retrieval (SIR) as follows. Let $b$ denote the research background (target problem), $i$ an inspiration (e.g., a publication), and $I$ the candidate literature pool.
Then SIR is expected to generate the ranking policy $P(i\mid b,I)$: given a fixed background $b$, rank candidates in $I$ so that publications useful for composing a hypothesis rise to the top.

\paragraph{Limitation.}
Existing SIR methods implicitly assume that the \emph{publication itself} is the retrieval object: each $i\in I$ is treated as an atomic document to be matched against $b$.
That assumption does not specify \emph{what} transferable content should be extracted from $i$, nor \emph{how} that content should depend on $b$.
Implementations therefore compare raw publication text to $b$ (or score usefulness only after generation), which conflates topical overlap with transferability. Moreover, it leaves no intermediate that explains why a retrieved publication can be useful in hypothesis generation for the target problem.
We instead argue that the retrieval object should be the \emph{transferable principle} extracted from the publication.
Human analogical transfer typically does not jump from a raw source instance to a target either: the source is first abstracted into a portable structural principle, then reinstantiated in the target~\citep{hope2017accelerating,boden2004creative}.
A classic illustration is biomimetic bridge design inspired by bone: a bone resists fracture while remaining light because a hard outer shell surrounds an internal web-like (trabecular) structure.
What transfers is not bone tissue or its exact morphology, but the abstract principle (a stiff envelope plus a sparse load-bearing lattice), which can be remapped onto materials and geometry suited to bridge construction.
Conditioning on the target matters: the same source (bone) admits many readings, and only the one needed for bridges is selected and remapped.
This is why we introduce \emph{target-conditioned abstraction} as an intermediate before ranking inspirations for SIR.

\paragraph{Multi-facet, multi-depth inspiration.}
A single scientific publication often contains multiple transferable ideas operating at different levels of abstraction.
Along the \emph{facet} axis, one may transfer a concrete component, an analysis method, an algorithm, or an entire problem-solving framework.
Along the \emph{depth} axis, the same idea can be stated in domain-specific terms or lifted toward a more domain-agnostic structural principle.
Different targets may need different facet--depth pairs from the same $i$.
Source-only abstraction therefore yields a set of candidate cores
\begin{equation}
\{a^{(1)},a^{(2)},\ldots,a^{(m)}\}=\mathrm{Abs}(i),
\end{equation}
but only one of them may be the reading that connects to $b$.
Useful SIR therefore requires \emph{selecting} the right abstraction for the target, not merely summarizing $i$.

\paragraph{Target-conditioned abstraction (TCA).}
To address the limitations of the previous formalization of the problem, we redefine SIR by introducing an explicit intermediate representation: a Reasoning--Abstraction (R/A) pair conditioned on both the candidate and the target,
\begin{equation}
\Phi(i,b)=(r_i,a_i),
\end{equation}
where $r_i$ is a reasoning trace that maps evidence in $i$ to $b$, and $a_i$ is the resulting transferable core (mechanism, hypothesis schema, bottleneck principle, etc.).
Transferability is scored from the conditioned R/A representation:
\begin{equation}
s_i=\tau(\Phi(i,b),b)\in[0,1].
\end{equation}
Ranking candidates by $s_i$ yields the retrieval policy
\begin{equation}
P(i\mid b,I)\;\propto\; \exp\!\big(\tau(\Phi(i,b),b)\big),
\end{equation}
over target-conditioned R/A pairs $\{\Phi(i,b):i\in I\}$.

\paragraph{Why this formalization is more useful.}
Compared with atomic publication ranking, TCA changes the object of retrieval from ``which publication is related to $b$?'' to ``which transferable core, when conditioned on $b$, best supports composing a hypothesis?''
This shift is valuable for three reasons.
First, it makes multi-facet / multi-depth ambiguity an explicit modeling target: $\Phi(\cdot,b)$ must choose the facet that the target needs.
Second, $(r_i,a_i)$ are inspectable intermediates that explain \emph{why} an inspiration is useful, which is critical for trusted use and for the downstream composition step $P(h\mid b,i)$.
Third, it yields a trainable objective for SIR itself: learn $\Phi$ and $\tau$ jointly so that the highly scored candidates are those whose conditioned abstraction is actually transferable, rather than those that merely share vocabulary with $b$.
Next we present our proposed method to address TCA, dubbed \textit{TCA-SIR}. TCA-SIR learns target-conditioned R/A generation together with a transfer score for ResearchBench-style retrieval.

%% file: sections/method.tex
\section{Method}

\begin{figure*}[t]
\centering
\includegraphics[width=0.92\textwidth]{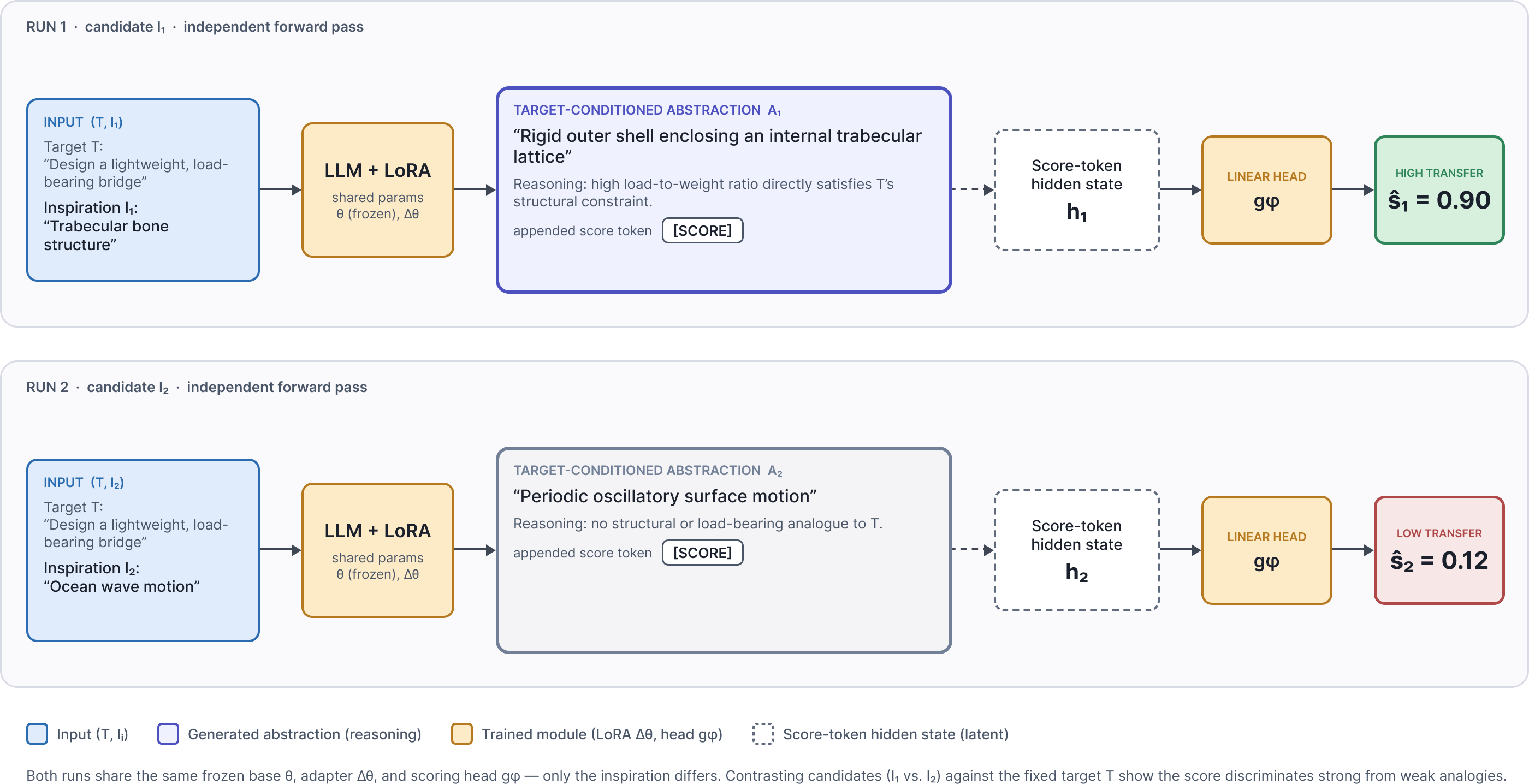}
\caption{Overview of TCA-SIR. Given a target problem and an inspiration candidate, an LLM fine-tuned with LoRA generates a Reasoning / target-conditioned abstraction and predicts a transferability score from the last-layer hidden state at a controller-inserted \texttt{Transfer score:} token via a linear score head. The figure shows two independent forward passes on the same target with two different inspirations as examples, yielding distinct abstractions and scores.}
\label{fig:method-overview}
\end{figure*}

\subsection{Overview}
TCA-SIR has two stages.
First, we build supervision for $\Phi$ and $\tau$: (i)~a \textbf{label-aware R/A teacher} that writes target-conditioned Reasoning/Abstraction text for each $(b,i)$; (ii)~a \textbf{generative judge} that filters those generations so only examples whose judged polarity matches the gold pair label are kept, as training-data quality control; and (iii)~a \textbf{transferability grader} that assigns a 5-level score that supervises continuous transfer strength.
Second, a LoRA student jointly learns to emit R/A ($\Phi$) and to predict that graded score from a readout-token hidden state ($\tau$), then ranks candidates by the predicted $s_i$.

\subsection{Training Data Construction}
Following the TCA formulation in the previous section, TCA-SIR jointly learns the abstraction function $\Phi$ and the transfer scoring function $\tau$.
Concretely, it maps a target problem $b$ and a candidate inspiration $i$ to a Reasoning--Abstraction (R/A) pair $(r_i,a_i)=\Phi(i,b)$ and a graded transfer score $s_i=\tau(r_i,a_i,b)\in[0,1]$.
Supervision for $\Phi$ and $\tau$ is built from ResearchBench~\citep{liu2026researchbench} training pairs with Llama-3.1-8B-Instruct as the sole teacher; therefore, label generation and later fine-tuning share one open checkpoint and avoid proprietary-model contamination.

\paragraph{Pair construction.}
We extract balanced positive/negative pairs from the ResearchBench train split: ground-truth inspirations are positives, and negatives preferentially use ResearchBench hard-tier candidates when available in the original dataset.

\paragraph{Label-aware R/A teacher.}
For each pair we generate target-conditioned reasoning and abstraction with \emph{separate} positive and negative teacher prompts that are conditioned on the gold label (transferable vs.\ scope-limited; Appendix~\ref{sec:supp-prompts}).
A frozen generative judge then assigns a continuous usefulness score to each generated abstraction and thresholds it into a binary decision (keep as positive if above the threshold, negative if below).
We retain a training example $(b,i,r_i,a_i)$ only when that decision agrees with the pair's gold label, and mismatched examples are discarded.

\paragraph{Graded transferability labels.}
Because binary positive/negative labels are coarse (even ground-truth inspirations vary in how strongly they help a target), a second LLM call assigns a 5-level mechanism-based transferability score $\{0.1,0.3,0.5,0.7,0.9\}$ under a fixed rubric that emphasizes adaptable mechanisms over topical overlap.
We use five evenly spaced grades rather than a binary label so the teacher can express intermediate transfer strength without requiring a fully continuous score that is hard to assign consistently, yielding nuanced supervision for the score head.
We score the target--candidate pair without showing the teacher R/A to the grader, so the grade reflects an independent transferability judgment rather than phrasing patterns of the abstractions that the score head could later recover as surface cues.
The grader does receive the gold binary label and tier as reference metadata (with an instruction not to copy them), which helps the LLM predict the scores with context of connections.
When the score distribution is unbalanced, an LLM rechecks adjacent scores (e.g., $0.7$ vs.\ $0.9$) and adjusts the scores.
Label-aware teacher prompts improve R/A supervision, but risk polarity-correlated phrasing that a score head could exploit as a shortcut. However, manual inspection found no fixed phrase-to-score templates across grades (Appendix~\ref{sec:supp-training-examples}).
The grader prompt appears with the teacher templates in Appendix~\ref{sec:supp-prompts}; graded-score distributions appear in Appendix~\ref{sec:supp-teacher-supervision}; we inspected samples and found Llama-generated R/A adequate for this supervision setting.

We intentionally use the same open-source LLM for teacher generation, judging, grading, and student initialization to avoid introducing external knowledge through a stronger or proprietary teacher. This design isolates the effect of the proposed target-conditioned abstraction representation rather than improvements arising from teacher distillation.

\subsection{Model Architecture}
Figure~\ref{fig:method-overview} depicts the proposed model architecture. The architecture has two modules: a LoRA-tuned generator that realizes $\Phi$ by emitting target-conditioned Reasoning/Abstraction text, and a linear score head that realizes $\tau$ by reading the hidden state after generation to predict transferability. TCA-SIR fine-tunes an LLM with LoRA~\citep{hu2022lora} on attention projections.
Given $(b,i)$, the model generates a text-anchored Reasoning and Abstraction sequence (headings \texttt{Reasoning:} / \texttt{Abstraction:}).
A lightweight linear score head then maps the last-layer hidden state at a controller-inserted \texttt{Transfer score:} readout token to a scalar transfer logit; at inference we apply a sigmoid to obtain $s_i=\tau(r_i,a_i,b)\in[0,1]$.
Thus scoring is not a separate decoder over R/A surface text: it reads the contextualized representation after the model has produced the full target-conditioned R/A sequence.

\subsection{Training Objective}
We jointly minimize a generation language-modeling loss on the R/A tokens and a mean-squared-error (MSE) loss between the predicted score and the graded transferability label:
\begin{equation}
\mathcal{L}=\lambda_{\mathrm{LM}}\,\mathcal{L}_{\mathrm{LM}}+\lambda_{s}\,\mathcal{L}_{\mathrm{MSE}}.
\end{equation}
The LM term trains $\Phi$ to emit useful, target-conditioned R/A; the MSE term trains $\tau$ to recover continuous transfer strength rather than a binary decision.
With $\lambda_{\mathrm{LM}}=\lambda_{s}=1$ on the reported checkpoint, joint optimization encourages the model to generate target-conditioned abstractions whose hidden representations are predictive of transferability.

%% file: sections/experiments.tex
\section{Experiments}

\input{tables/generated/main_comparison}

\subsection{Dataset}
We evaluate on the \textbf{Inspiration Retrieval} task from ResearchBench~\citep{liu2026researchbench}.
ResearchBench covers 12 scientific domains and is built from papers published in 2024 onward to reduce overlap with LLM pretraining corpora.
For each target, the input is a research question and a background survey. The system must rank a fixed pool of 75 candidate papers, each represented by title and abstract only.
Every pool contains 2--3 gold inspirations that contributed to the original paper's hypothesis, together with hard negatives sampled at multiple topical distances from the target.
The ranking goal is to recover those gold inspirations within the pool.
We construct domain-balanced holdouts with an equal number of targets per domain: \textbf{test240} uses 20 targets/domain (headline results), and \textbf{val120} uses 10 targets/domain.
Training uses all remaining complete-pool targets after the test and validation holdouts (domain-balanced remainder; not a fixed per-domain train quota).
Sensitivity sweeps use \textbf{pilot12} (one target per domain from val120) to avoid test leakage.
Train, validation, and test \emph{target} sets are enforced to be disjoint in the split pipeline, and pre-training contamination checks verify that training pairs never include held-out evaluation targets.

\subsection{Evaluation metrics}
Following ResearchBench~\citep{liu2026researchbench}, we report HitRate@top4\% and HitRate@top20\% as the fraction of gold inspirations recovered in the top $3/75$ or $15/75$ of each pool, averaged over targets.
We also report \textbf{MRR} (reciprocal rank of the first gold inspiration) and \textbf{NDCG@3} over the top three positions (mean $\pm$ bootstrap 95\% CI half-width); the NDCG discount formula appears in Appendix~\ref{sec:supp-implementation}.

\subsection{Baselines}
All methods use Llama-3.1-8B-Instruct at temperature $0$ under the same two-round retrieval protocol (75 candidates $\rightarrow$ 5 groups of 15 $\rightarrow$ shortlist $\rightarrow$ top 3).
We evaluate on a single open backbone so that gains can be attributed to the retrieval formulation rather than to differences in underlying LLM capacity.
We choose Llama-3.1-8B-Instruct because its pretraining cutoff (December 2023) precedes ResearchBench's source papers (2024 onward), reducing the risk of pretraining contamination.
We organize baselines into three groups: \emph{direct retrieval} methods that rank papers without an explicit abstraction intermediate (strong untrained references); \emph{prior abstraction} methods adapted from related SIR work; and \emph{untrained TCA controls} that share our Reasoning/Abstraction interface but without learning.

\paragraph{Direct retrieval.}
\emph{Direct LLM (MOOSE-Chem)} applies the original MOOSE-Chem prompt~\citep{yang2025moose} to select the most useful inspirations from each candidate group.
\emph{Direct LLM pair score} scores each target--candidate pair with the same five-level transferability rubric used for TCA-SIR supervision, without generating an abstraction.

\paragraph{Prior abstraction methods.}
\emph{Idea-Catalyst-style} extracts domain-agnostic core problems for the target and candidates~\citep{kargupta2026sparking}, then applies listwise selection over those representations.
\emph{Gen-level retrieval} generates four abstraction levels for the target and candidate~\citep{gu2024llms} and ranks candidates using cosine similarity between Llama hidden-state embeddings.
The \emph{aligned} variant compares corresponding levels, whereas the \emph{max} variant uses the highest cross-level similarity.

\paragraph{Untrained TCA controls.}
\emph{SourceAbs} generates a target-agnostic abstraction of each candidate and then scores its usefulness for the target.
\emph{Prompt-TCA} uses the same target-conditioned Reasoning/Abstraction interface as TCA-SIR, but without fine-tuning or a learned score head.

\subsection{Implementation details}
TCA-SIR fine-tunes with LoRA~\citep{hu2022lora} ($r{=}8$, $\alpha{=}16$ on \texttt{q\_proj}/\texttt{v\_proj}) for 3 epochs at learning rate $1{\times}10^{-5}$ with score-loss weight $1.0$; sensitivity on pilot12 is in Appendix~\ref{sec:supp-sensitivity}.
Hardware and full run configs appear in Appendix~\ref{sec:supp-implementation}.

%% file: tables/generated/main_comparison.tex
\begin{table*}[!t]
\centering
\small
\begin{tabular}{lcccc}
\toprule
Method & HitRate@top4\% & HitRate@top20\% & MRR & NDCG@3 \\
\midrule
SourceAbs & $0.189 \pm 0.035$ & $0.558 \pm 0.043$ & $0.245 \pm 0.047$ & $0.173 \pm 0.033$ \\
Prompt-TCA & $0.219 \pm 0.034$ & $0.708 \pm 0.041$ & $0.290 \pm 0.048$ & $0.203 \pm 0.033$ \\
Gen-level (aligned) & $0.263 \pm 0.039$ & $0.694 \pm 0.041$ & $0.361 \pm 0.053$ & $0.258 \pm 0.040$ \\
Gen-level (max) & $0.289 \pm 0.042$ & $0.690 \pm 0.041$ & $0.376 \pm 0.054$ & $0.279 \pm 0.041$ \\
Idea-Catalyst & $0.299 \pm 0.041$ & $0.655 \pm 0.044$ & $0.414 \pm 0.055$ & $0.296 \pm 0.042$ \\
Direct LLM & $0.371 \pm 0.042$ & $0.750 \pm 0.037$ & $0.511 \pm 0.055$ & $0.371 \pm 0.042$ \\
Direct LLM pair score & $0.379 \pm 0.044$ & $0.808 \pm 0.036$ & $0.483 \pm 0.055$ & $0.371 \pm 0.044$ \\
TCA-SIR & {\boldmath$0.481 \pm 0.045$} & {\boldmath$0.850 \pm 0.031$} & {\boldmath$0.601 \pm 0.054$} & {\boldmath$0.472 \pm 0.045$} \\
\bottomrule
\end{tabular}
\caption{Test240 scientific inspiration retrieval. Cells report mean $\pm$ half-width of a bootstrap 95\% CI over targets. Rows are ordered from simpler baselines to our final model (\textbf{TCA-SIR}, lr~$1{\times}10^{-5}$). Best value in each column is bold.}
\label{tab:main-comparison}
\end{table*}

%% file: sections/results.tex
\section{Results}

\subsection{Main comparison}

Table~\ref{tab:main-comparison} summarizes HitRate at top-4\% / top-20\%, MRR, and NDCG@3 on test240.
\textbf{TCA-SIR} is best on all four metrics.
Among untrained methods, the Direct LLM baselines are strongest: MOOSE-Chem listwise screening and Direct LLM pair score both outperform every prompt-based abstraction method.
Notably, all untrained abstraction methods underperform Direct LLM, whereas trained TCA-SIR substantially exceeds it, indicating that abstraction becomes effective only when learned for the retrieval objective.

Idea-Catalyst is the best \emph{untrained abstraction} baseline, but still trails Direct LLM.
It reuses the MOOSE-Chem listwise harness after replacing each publication with a one-sentence, domain-agnostic core problem; those cores often wash out mechanism detail (Appendix~\ref{sec:supp-additional-results}); therefore, listwise selection loses surface evidence that Direct LLM still uses.
Gen-level ranks next: an L1--L4 ladder retains more structure than a single abstraction, but high levels collapse into vague language and mismatched level pairs can bury the true inspiration (Appendix~\ref{sec:supp-additional-results}, Cases~G--I).
Prompt-TCA outperforms SourceAbs because its abstraction is target-conditioned, but zero-shot R/A often hedges and underranks true inspirations (e.g., ranks 7--10 on Cases~G--I).
SourceAbs is the weakest: target-agnostic paraphrase collapses one inspiration's many facets into a single aspect that need not align with the target.
Taken together, inspiration abstraction is \emph{multi-faceted} and \emph{multi-depth}: choosing the wrong facet or depth loses the SIR signal, and untrained prompting alone does not reliably select it.

\subsection{Qualitative target-conditioned abstraction}

\input{tables/qualitative_main_abstractions}

Beyond ranking metrics, we inspect whether TCA-SIR extracts a \emph{target-relevant mechanism} rather than topical similarity.
Table~\ref{tab:qual-main-abstractions} compares TCA-SIR and Prompt-TCA on two cases chosen so the two methods abstract the inspiration differently (ranks: TCA-SIR / Prompt-TCA).
\begin{itemize}
\item \textbf{Example~A} (battery materials).
TCA-SIR recovers the inspiration's \emph{specific chemical lever} (amide / ethylenediamine groups that bind unwanted intermediates) and retargets it into electrode design; Prompt-TCA collapses to a near-template claim (``use functionalized carbon'') that any related paper could have produced (ranks 1 / 3).
\item \textbf{Example~B} (space instrumentation).
TCA-SIR selects the inspiration's \emph{performance bottleneck} (reduce heat leak so cryogen lasts long enough) rather than restating system inventory; Prompt-TCA mainly lists a co-present hardware fact (hybrid coolers) without committing to that bottleneck-solving principle (both rank 1).
\end{itemize}
Example~C (hypothesis-schema / environmental economics) appears in Appendix~\ref{sec:supp-qualitative}.
Overall, TCA-SIR recovers target-relevant portable cores, whereas Prompt-TCA often produces generic or secondary observations.

\subsection{Ablation Study}

\input{tables/generated/ablation_ladder}

Table~\ref{tab:ablation-ladder} isolates the stack from source-only prompting to full TCA-SIR.
Target conditioning helps modestly (Prompt-TCA $>$ SourceAbs), but the gain is small: without training, prompting for ``an abstraction'' often misses the facet the target needs (Table~\ref{tab:qual-main-abstractions}; SourceAbs in Appendix~\ref{sec:supp-qualitative}).

The decisive jump is supervised learning.
\emph{TCA-SIR (binary BCE)} keeps R/A generation and the score head, but trains the head with binary BCE instead of graded scores.
\emph{TCA-SIR (score-only)} disables R/A generation ($\lambda_{\mathrm{LM}}=0$) and trains only the graded score head.
\emph{TCA-SIR (no reasoning)} generates Abstraction and a graded score but drops Reasoning.
Replacing Prompt-TCA with binary BCE nearly doubles HitRate@top4\%, so most of the gain comes from learning rather than the TCA prompt alone.
Graded scores then beat binary labels, indicating that continuous transfer strength is a better training target than a coarse positive/negative split, since not all ground-truth inspirations are equally useful.
Score-only stays close to full TCA-SIR, but the drop shows that generating abstractions also helps ranking, not only interpretability.
Omitting Reasoning still hurts, suggesting the trace is part of applying an abstraction to a new target.
Across this ladder, trained TCA-SIR variants exceed Direct LLM while every untrained abstraction method falls below it (Table~\ref{tab:main-comparison}).
Prompt-only intermediates are an inductive bias that needs learning signal; full TCA-SIR recovers the portable core illustrated qualitatively above.

%% file: tables/qualitative_main_abstractions.tex
\begin{table*}[t]
\centering
\small
\setlength{\tabcolsep}{1mm}
\begin{tabular}{@{}p{0.08\textwidth}p{0.28\textwidth}p{0.28\textwidth}p{0.28\textwidth}@{}}
\toprule
& \textbf{Target / inspiration} & \textbf{TCA-SIR} & \textbf{Prompt-TCA} \\
\midrule
A\newline
{\small ranks 1 / 3}
&
\textit{How can we design a high-performance lithium-sulfur battery electrode material that simultaneously addresses the problems of low sulfur loading, polysulfide shuttle effect, and sluggish polysulfide redox kinetics, leading to improved capacity, cycling stability, and rate capability?}
\par
Inspiration: \emph{Strong adsorption of lithium polysulfides on ethylenediamine-functionalized carbon fiber paper interlayer providing excellent capacity retention of lithium-sulfur batteries}
&
A high-performance lithium-sulfur battery electrode material can be designed by incorporating a functionalized interlayer with amide groups that strongly adsorb lithium polysulfides, thereby reducing the polysulfide shuttle effect and improving capacity retention, cycling stability, and rate capability. This approach can be achieved by modifying the surface of the electrode material with amide-containing functional groups, which can effectively trap high-order polysulfides and facilitate their conversion to solid lithium sulfides, thereby addressing the challenges of low sulfur loading, polysulfide shuttle effect, and sluggish polysulfide redox kinetics.
&
A key principle that can be abstracted from the inspiration paper is the use of functionalized carbon materials to enhance the adsorption of lithium polysulfides. This principle can be applied to the target problem by designing high-performance lithium-sulfur battery electrode materials that incorporate functional groups capable of strong interactions with lithium polysulfides. By leveraging this principle, researchers can develop materials that effectively mitigate the polysulfide shuttle effect and improve the overall performance of lithium-sulfur batteries.
\\
\midrule
B\newline
{\small ranks 1 / 1}
&
\textit{How can the design and performance of a helium Dewar be improved for the Resolve instrument onboard the XRISM satellite to meet the stringent cooling requirements for an X-ray microcalorimeter array, ensuring a thermal interface below 1.5 K and a helium lifetime of over 3 years in orbit?}
\par
Inspiration: \emph{Flight model performance test results of a helium dewar for the soft X--ray spectrometer onboard ASTRO-H}
&
A key principle for improving the design and performance of the helium Dewar for the Resolve instrument is to incorporate innovative design improvements and rigorous testing to meet the stringent cooling requirements, such as reducing the heat load to the helium tank and minimizing heat transfer through multiple stages of insulation and cooling, thereby ensuring a thermal interface below 1.5 K and a helium lifetime of over 3 years in orbit.
&
A transferable mechanism principle from the inspiration is the incorporation of a hybrid cooling system that includes both liquid helium and mechanical coolers, which can be adapted to the target problem to minimize heat transfer and maximize efficiency. This principle can be applied to the design of the helium Dewar for the Resolve instrument, potentially leading to improved thermal performance and a longer helium lifetime.
\\
\bottomrule
\end{tabular}
\caption{Model-generated abstractions for Examples~A--B (ranks: TCA-SIR / Prompt-TCA).}
\label{tab:qual-main-abstractions}
\end{table*}

%% file: tables/generated/ablation_ladder.tex
\begin{table}[t]
\centering
\small
\setlength{\tabcolsep}{1mm}
\begin{tabular}{@{}lcc@{}}
\toprule
Method & HitRate@top4\% & MRR \\
\midrule
SourceAbs & $0.189 \pm 0.035$ & $0.245 \pm 0.047$ \\
Prompt-TCA & $0.219 \pm 0.034$ & $0.290 \pm 0.048$ \\
TCA-SIR (binary BCE) & $0.417 \pm 0.042$ & $0.545 \pm 0.054$ \\
TCA-SIR (score-only) & $0.443 \pm 0.048$ & $0.535 \pm 0.056$ \\
TCA-SIR (no reasoning) & $0.446 \pm 0.047$ & $0.544 \pm 0.056$ \\
TCA-SIR & {\boldmath$0.481 \pm 0.045$} & {\boldmath$0.601 \pm 0.054$} \\
\bottomrule
\end{tabular}
\caption{Ablation ladder from source-only prompting to full TCA-SIR (lr~$1{\times}10^{-5}$) on test240 (mean $\pm$ bootstrap 95\% CI half-width). Best value in each column is bold.}
\label{tab:ablation-ladder}
\end{table}

%% file: sections/conclusion.tex
\section{Conclusions and Future Work}

Scientific inspiration retrieval (SIR) is a bottleneck for LLM-based hypothesis generation. Useful inspirations are often topically remote, yet carry a transferable principle for the target.
Drawing on how humans abstract and reinstantiate remote analogies, we redefined SIR as TCA (Target-Conditioned Abstraction), where the retrieval object is a transferable principle extracted from a candidate and conditioned on the target rather than the publication as an atomic document.
We proposed TCA-SIR, which jointly learns target-conditioned abstractions and a transferability score for ranking candidates.
On ResearchBench, TCA-SIR improves HitRate and ranking metrics over direct LLM prompting and abstraction-based baselines, and qualitative analysis shows that learned abstractions recover target-relevant mechanisms rather than generic topical similarities.
A practical limitation is computational cost. Because each abstraction is target-conditioned, TCA-SIR must generate an R/A intermediate for every target-candidate pair rather than a single source-side summary, making large pools expensive at inference.
Future work will explore more efficient ways to scale this approach.

\section*{Use of AI Tools}
Generative AI tools were used to assist with language editing, code development and debugging, and figure generation and refinement. The authors reviewed and verified all AI-assisted outputs and take full responsibility for the manuscript and its supporting materials.

%% file: appendix/overview.tex
\section{Overview}
\label{sec:supp-overview}

This appendix provides extended experimental results for the main text.
Naming follows the main text: \textbf{TCA-SIR} (trained target-conditioned abstraction), \textbf{Prompt-TCA} (prompted target-conditioned abstraction), and \textbf{SourceAbs} (source-only abstraction).
Qualitative tables reprint full generations for main-text Examples~A--B, present Example~C (hypothesis-schema case) here, and add same-inspiration / cross-task cases, plus Gen-level L1--L4 and Idea-Catalyst core-problem intermediates that help explain those baselines' weaker retrieval.
We also include teacher-supervision quality checks, the teacher and grader prompts used to construct TCA-SIR training supervision, and representative positive/negative training R/A examples.

%% file: appendix/additional_results.tex
\section{Additional Results}
\label{sec:supp-additional-results}

\subsection{Sensitivity (pilot12)}
\label{sec:supp-sensitivity}
Table~\ref{tab:sensitivity-pilot12} and Figures~\ref{fig:sens-lr}--\ref{fig:sens-instruction} summarize learning-rate, score-loss weight, and instruction-detail sweeps on pilot12.
Within the explored ranges, learning rate $1{\times}10^{-5}$ and score-loss weight~$1.0$ (our reference setting) perform best; verbose instructions underperform the standard/minimal prompts.

\input{tables/generated/sensitivity_pilot12}

\begin{center}
\includegraphics[width=0.88\columnwidth]{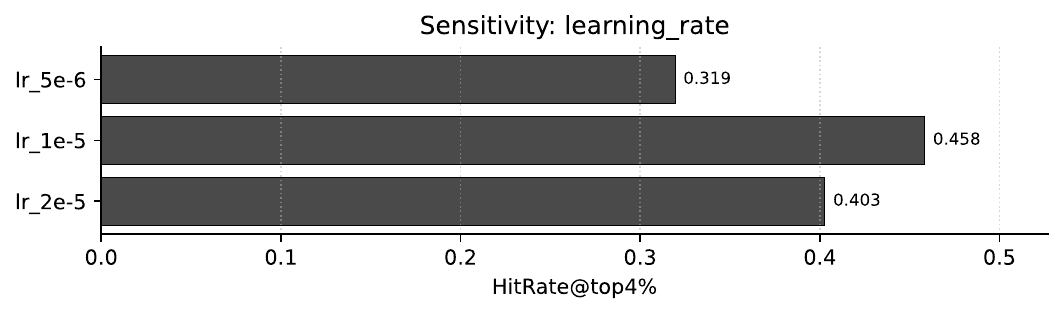}
\captionof{figure}{Learning-rate sensitivity on pilot12 (HitRate@top4\%).}
\label{fig:sens-lr}
\end{center}

\begin{center}
\includegraphics[width=0.88\columnwidth]{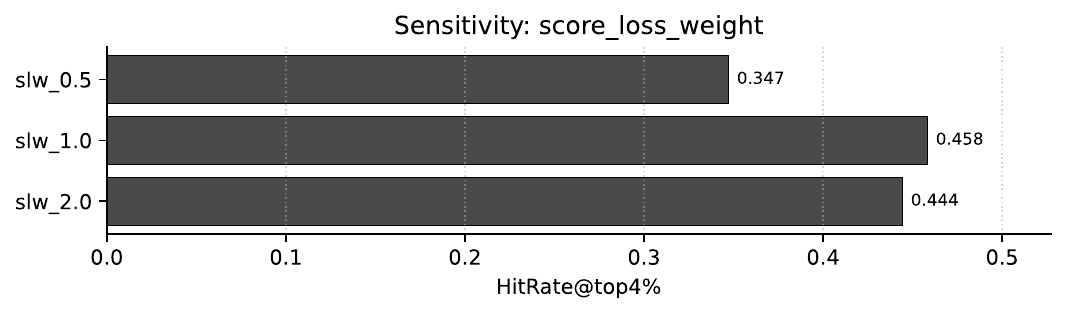}
\captionof{figure}{Score-loss weight sensitivity on pilot12 (HitRate@top4\%).}
\label{fig:sens-slw}
\end{center}

\begin{center}
\includegraphics[width=0.88\columnwidth]{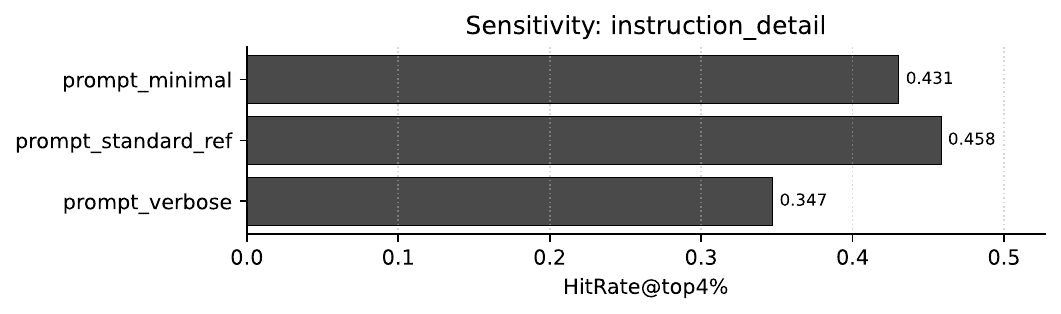}
\captionof{figure}{Instruction-detail sensitivity on pilot12 (HitRate@top4\%).}
\label{fig:sens-instruction}
\end{center}

\subsection{Qualitative examples (extended)}
\label{sec:supp-qualitative}
Table~\ref{tab:qual-supp-example-a} presents Example~C, an environmental-economics case moved out of the main text.
TCA-SIR recovers a \emph{structured hypothesis} from the inspiration (fiscal policy shapes an Environmental Kuznets Curve) and remaps it to a new outcome variable (load capacity factor) and setting (BRICS).
Prompt-TCA only offers a soft advice slogan (``consider fiscal policy / greener energy'') without that reusable model (ranks 2 / 5).
This case illustrates recovery of a transferable hypothesis schema rather than a concrete chemical lever or performance bottleneck.
Table~\ref{tab:qual-supp-sourceabs-abc} reports SourceAbs generations for main-text Examples~A--B and Appendix Example~C.
Table~\ref{tab:qual-supp-former-abc} reprints earlier qualitative cases (Examples~G--I: queuing/edge offloading, pericyte atlas, PINN shocks), where TCA-SIR and Prompt-TCA often share a mechanism family and Prompt-TCA mainly under-ranks.
Table~\ref{tab:qual-supp-extra} adds Examples~D--F (space robotics, disulfide/disulfidptosis, MXene/SiRF).
Table~\ref{tab:qual-supp-paa} holds the inspiration fixed (anionic polyacrylamide adsorption study) and varies the target: TCA-SIR emphasizes experimental--computational characterization for a specific PAA@TiO$_2$ nanocomposite target, but polymer--nanoparticle integration for a generic cost-efficient adsorbent target.
This is preliminary evidence of target-sensitive facet selection; the two targets remain closely related, so we do not treat them as a conclusive cross-goal counterfactual.

\input{tables/qualitative_supp_example_a}
\input{tables/qualitative_supp_sourceabs_abc}
\input{tables/qualitative_supp_former_abc}
\input{tables/qualitative_supp_extra}
\input{tables/qualitative_supp_paa}

\paragraph{Score--generation disagreement.}
TCA-SIR's transfer score and generated abstraction are not perfectly aligned.
In some held-out cases the model ranks a ground-truth inspiration highly while the generated text is cautious or scope-limited (e.g., nonlinear programming for satellite orbit prediction; economic-growth/environment work for renewable-energy spillovers; zinc-complex synthesis for antileukemia complexes).
A high retrieval score should therefore not always be read as evidence of a faithful actionable abstraction; improving score--generation consistency is future work.

\subsection{Why Gen-level and Idea-Catalyst underperform}
\label{sec:supp-genlevel-idea}
The main-paper comparison shows that Gen-level and Idea-Catalyst lag Direct LLM and TCA-SIR.
Tables~\ref{tab:qual-supp-genlevel} and~\ref{tab:qual-supp-idea-catalyst} inspect intermediate artifacts on Examples~G--I (SourceAbs for Examples~A--C appears in Table~\ref{tab:qual-supp-sourceabs-abc}).
Gen-level L3--L4 texts and Idea-Catalyst core problems both tend to erase domain-specific mechanism details in favor of generic ``complex systems'' language, which weakens ranking for scientific inspiration transfer.

\input{tables/qualitative_supp_genlevel}
\input{tables/qualitative_supp_idea_catalyst}
\FloatBarrier

\subsection{Per-discipline breakdown}
\label{sec:supp-per-discipline}
Figure~\ref{fig:per-discipline} reports HitRate@top4\% by ResearchBench domain for headline methods.
Aggregate test240 gains for TCA-SIR are not driven by a single discipline; relative orderings remain broadly stable across domains.

\begin{center}
\includegraphics[width=\columnwidth]{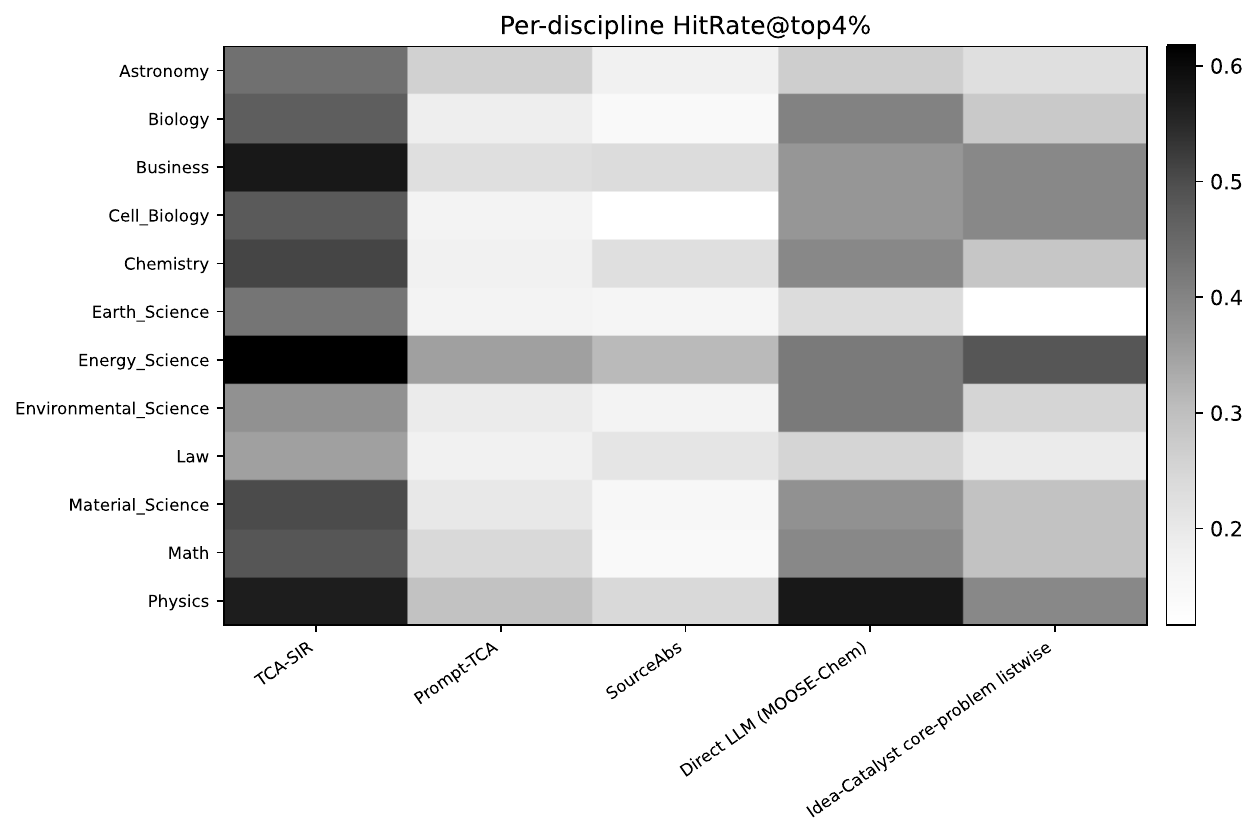}
\captionof{figure}{Per-discipline HitRate@top4\% heatmap on test240 for headline methods.}
\label{fig:per-discipline}
\end{center}

\subsection{Training dynamics and extra ranking views}
\label{sec:supp-training-dynamics}
Figure~\ref{fig:train-val-loss} shows train loss components and validation trends for the TCA-SIR run used in the main comparison (lr~$1{\times}10^{-5}$).
Figures~\ref{fig:trained-vs-untrained}--\ref{fig:ablations-hitrate-supp} give extra ranking views: trained vs.\ untrained HitRate@top4\%, overall MRR, overall HitRate@top4\%, and ablation / Prompt-TCA / SourceAbs HitRate@top4\%.

\begin{center}
\includegraphics[width=0.78\columnwidth]{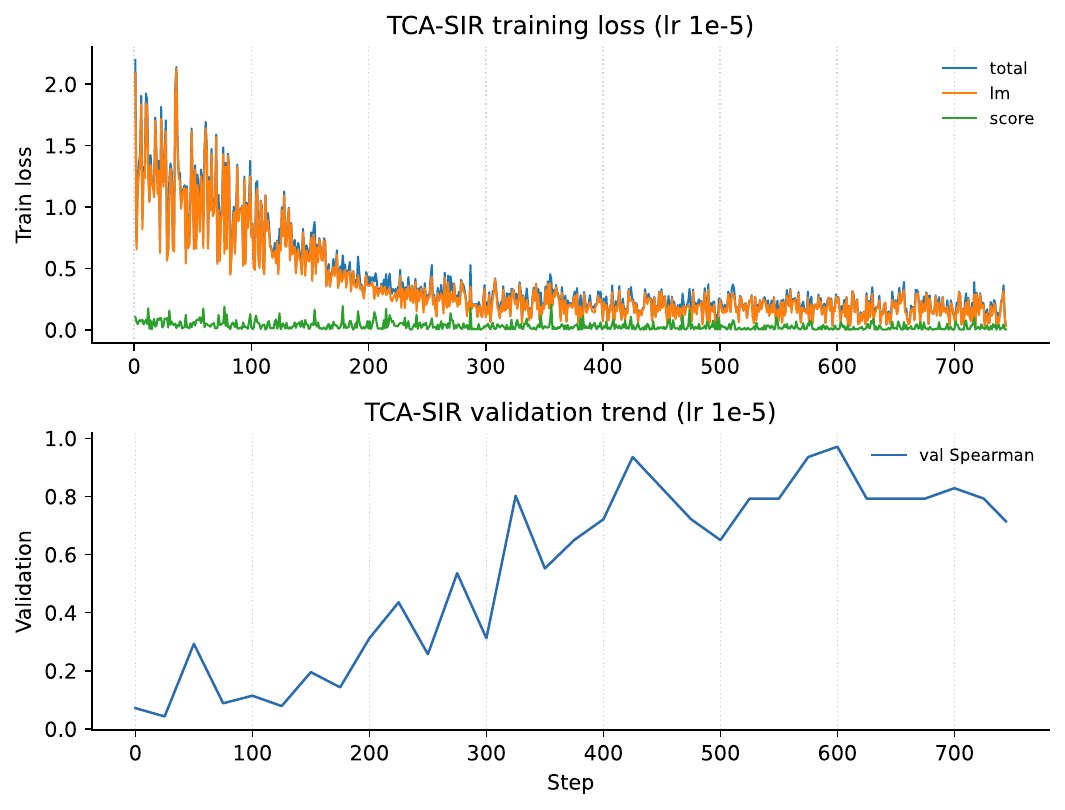}
\captionof{figure}{TCA-SIR training loss and validation trend.}
\label{fig:train-val-loss}
\end{center}

\begin{center}
\includegraphics[width=0.78\columnwidth]{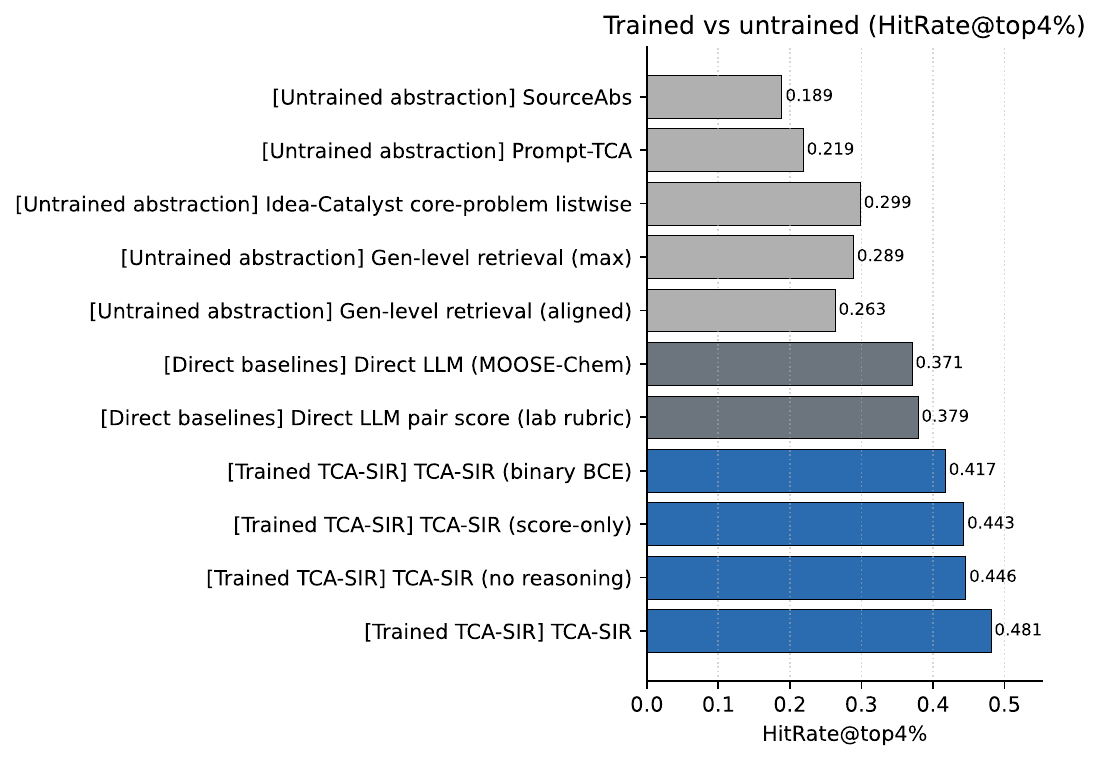}
\captionof{figure}{Trained vs.\ untrained HitRate@top4\% on test240.}
\label{fig:trained-vs-untrained}
\end{center}

\begin{center}
\includegraphics[width=0.78\columnwidth]{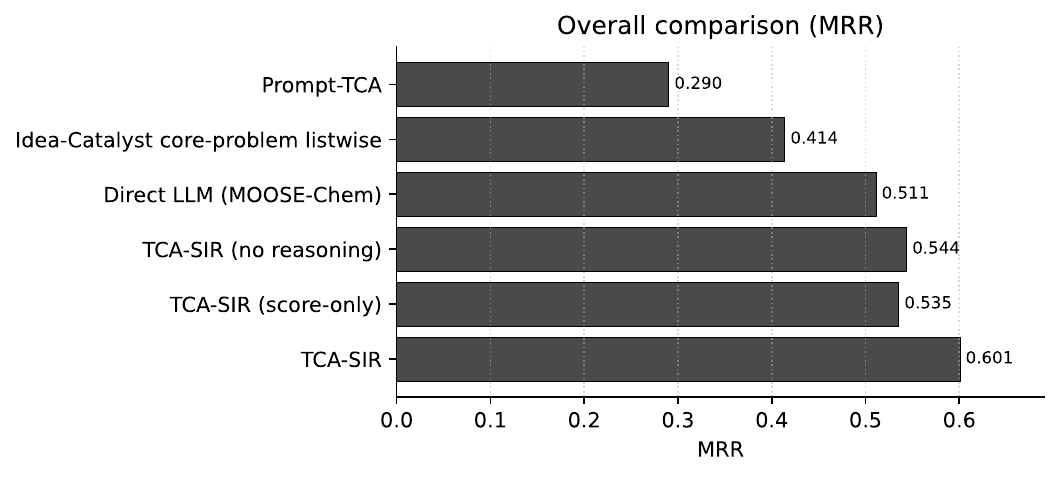}
\captionof{figure}{Overall MRR ranking on test240.}
\label{fig:overall-mrr-supp}
\end{center}

\begin{center}
\includegraphics[width=0.78\columnwidth]{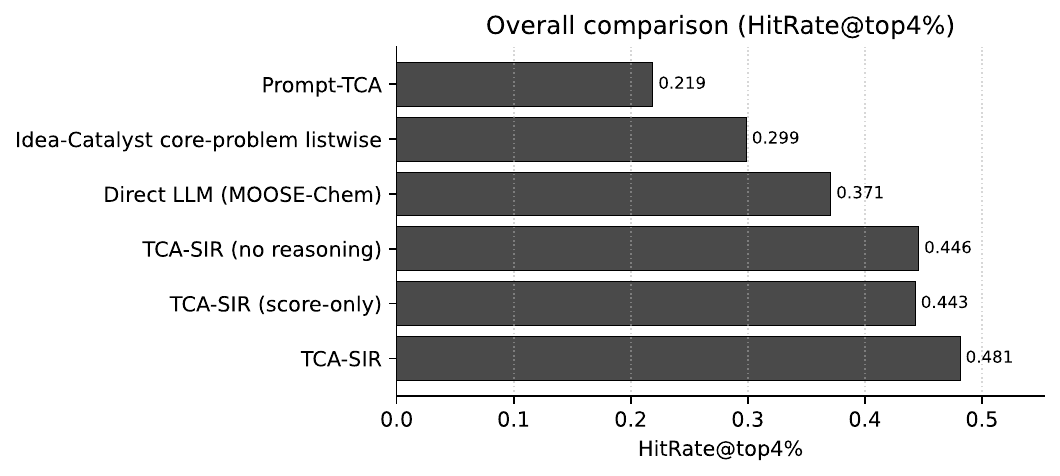}
\captionof{figure}{Overall HitRate@top4\% ranking on test240.}
\label{fig:overall-hitrate4-supp}
\end{center}

\begin{center}
\includegraphics[width=0.78\columnwidth]{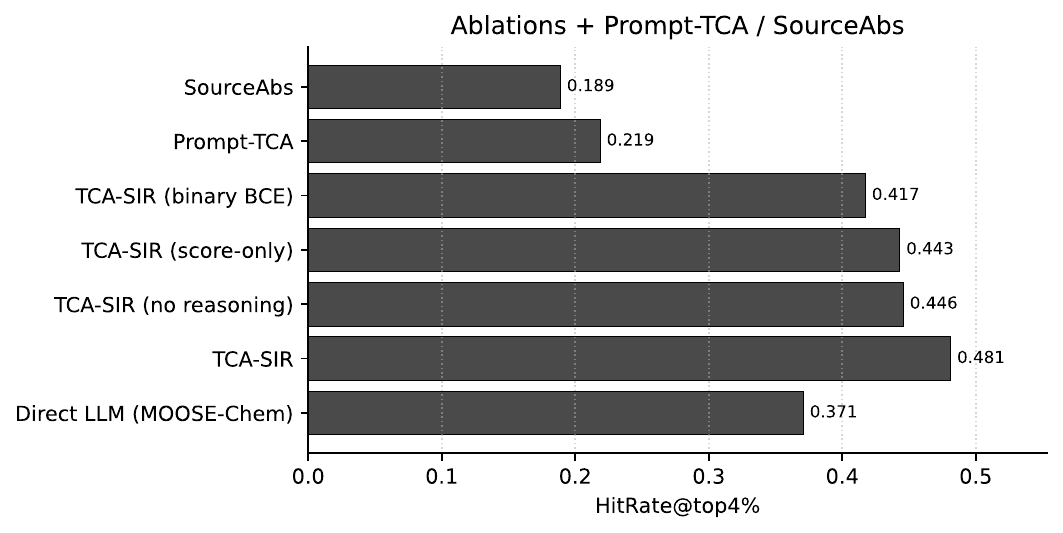}
\captionof{figure}{Ablation / Prompt-TCA / SourceAbs HitRate@top4\%.}
\label{fig:ablations-hitrate-supp}
\end{center}

%% file: tables/generated/sensitivity_pilot12.tex
\begin{center}
\small
\begin{tabular}{llc}
\toprule
Axis & Setting & HitRate@top4\% \\
\midrule
Learning rate & $5{\times}10^{-6}$ & 0.319 \\
Learning rate & $1{\times}10^{-5}$ & {\bf 0.458} \\
Learning rate & $2{\times}10^{-5}$ & 0.403 \\
\midrule
Score-loss weight & 0.5 & 0.347 \\
Score-loss weight & 1.0 & {\bf 0.458} \\
Score-loss weight & 2.0 & 0.444 \\
\midrule
Instruction detail & minimal & 0.431 \\
Instruction detail & standard (ref.) & {\bf 0.458} \\
Instruction detail & verbose & 0.347 \\
\bottomrule
\end{tabular}
\captionof{table}{Sensitivity on pilot12 (12-target subset of val120). HitRate@top4\% for learning rate, score-loss weight, and instruction-detail sweeps. Best value within each axis is bold.}
\label{tab:sensitivity-pilot12}
\end{center}

%% file: tables/qualitative_supp_example_a.tex
\begin{table*}[t]
\centering
\small
\setlength{\tabcolsep}{1mm}
\begin{tabular}{@{}p{0.08\textwidth}p{0.28\textwidth}p{0.28\textwidth}p{0.28\textwidth}@{}}
\toprule
& \textbf{Target / inspiration} & \textbf{TCA-SIR} & \textbf{Prompt-TCA} \\
\midrule
C\newline
{\small ranks 2 / 5}
&
\textit{How does fiscal policy (specifically, government expenditure and taxation revenue), along with renewable and non-renewable energy consumption and economic growth, impact the load capacity factor (LCF), a comprehensive measure of environmental quality, in BRICS nations (Brazil, Russia, India, China, and South Africa), considering potential asymmetric effects?}
\par
Inspiration: \emph{Testing the role of fiscal policy in the environmental degradation: The case of Turkey}
&
A fiscal policy-induced Environmental Kuznets Curve (EKC) hypothesis can be applied to BRICS nations, where government expenditure and taxation revenue influence the load capacity factor (LCF), a comprehensive measure of environmental quality, by reducing carbon emissions and promoting sustainable economic growth, particularly when fiscal policies support energy efficiency and alternative energy systems.
&
A transferable mechanism principle from the inspiration paper is the importance of integrating fiscal policy analysis into environmental studies, particularly in the context of emerging economies. This principle can be adapted to the target problem by emphasizing the need for policymakers in BRICS nations to consider the environmental implications of their fiscal decisions when designing policies to promote economic growth and reduce environmental degradation. Specifically, the inspiration paper's findings suggest that policymakers should focus on enhancing fiscal policies that support energy efficiency and promote alternative energy systems, which can be applied to the target problem by highlighting the need for BRICS nations to transition towards greener energy practices and reduce their reliance on non-renewable energy sources.
\\
\bottomrule
\end{tabular}
\caption{Environmental-economics case (Example~C). Model-generated abstractions (ranks: TCA-SIR / Prompt-TCA).
SourceAbs for this case is in Table~\ref{tab:qual-supp-sourceabs-abc}.}
\label{tab:qual-supp-example-a}
\end{table*}

%% file: tables/qualitative_supp_sourceabs_abc.tex
\begin{table*}[t]
\centering
\small
\setlength{\tabcolsep}{1mm}
\begin{tabular}{@{}p{0.08\textwidth}p{0.28\textwidth}p{0.55\textwidth}@{}}
\toprule
& \textbf{Target / inspiration} & \textbf{SourceAbs} \\
\midrule
A\newline
{\small rank 1}
&
\textit{How can we design a high-performance lithium-sulfur battery electrode material that simultaneously addresses the problems of low sulfur loading, polysulfide shuttle effect, and sluggish polysulfide redox kinetics, leading to improved capacity, cycling stability, and rate capability?}
\par
Inspiration: \emph{Strong adsorption of lithium polysulfides on ethylenediamine-functionalized carbon fiber paper interlayer providing excellent capacity retention of lithium-sulfur batteries}
&
Designing and implementing tailored materials with specific functional groups to enhance chemical interactions with target species can significantly improve the performance of energy storage systems by mitigating key challenges such as capacity degradation and shuttle effects.
\\
\midrule
B\newline
{\small rank 3}
&
\textit{How can the design and performance of a helium Dewar be improved for the Resolve instrument onboard the XRISM satellite to meet the stringent cooling requirements for an X-ray microcalorimeter array, ensuring a thermal interface below 1.5 K and a helium lifetime of over 3 years in orbit?}
\par
Inspiration: \emph{Flight model performance test results of a helium dewar for the soft X--ray spectrometer onboard ASTRO-H}
&
Design and testing of complex systems under stringent requirements can be achieved through a combination of rigorous engineering, innovative design improvements, and comprehensive testing. This approach enables the identification and mitigation of potential failure modes, ultimately leading to the development of reliable and efficient systems that meet or exceed performance expectations.
\\
\midrule
C\newline
{\small rank 2}
&
\textit{How does fiscal policy (specifically, government expenditure and taxation revenue), along with renewable and non-renewable energy consumption and economic growth, impact the load capacity factor (LCF), a comprehensive measure of environmental quality, in BRICS nations (Brazil, Russia, India, China, and South Africa), considering potential asymmetric effects?}
\par
Inspiration: \emph{Testing the role of fiscal policy in the environmental degradation: The case of Turkey}
&
Integrate policy analysis into studies of complex systems to uncover the interplay between policy variables and system outcomes, and to identify potential policy levers that can be manipulated to achieve desired system behaviors. This involves analyzing the relationships between policy variables, system indicators, and outcomes, and using statistical and econometric techniques to test hypotheses and identify causal relationships. By doing so, researchers can develop a deeper understanding of the dynamics of complex systems and identify effective policy strategies that can be used to manage system outcomes and achieve desired goals.
\\
\bottomrule
\end{tabular}
\caption{SourceAbs generations for Examples~A--C (group rank of the ground-truth inspiration).
Examples~A--B appear in the main paper; Example~C is discussed in this supplement.}
\label{tab:qual-supp-sourceabs-abc}
\end{table*}

%% file: tables/qualitative_supp_former_abc.tex
\begin{table*}[t]
\centering
\small
\setlength{\tabcolsep}{1mm}
\begin{tabular}{@{}p{0.08\textwidth}p{0.28\textwidth}p{0.28\textwidth}p{0.28\textwidth}@{}}
\toprule
& \textbf{Target / inspiration} & \textbf{TCA-SIR} & \textbf{Prompt-TCA} \\
\midrule
G\newline
{\small ranks 1 / 7}
&
\textit{How can we effectively manage computation offloading in a multi-mobile user (MU), multi-edge server (ES) environment where ES resources are heterogeneous and finite, considering the need to minimize both MU time and energy consumption while maximizing ES resource utilization?}
\par
Inspiration: \emph{A queuing theory model for cloud computing}
&
A queuing theory-based approach can be applied to model the computation offloading process in a multi-user, multi-edge server environment, where edge servers are treated as queuing systems with finite capacities, and task allocation is optimized to minimize waiting times and maximize resource utilization, thereby balancing MU time and energy consumption while maximizing ES resource utilization.
&
A queuing theory-based approach can be adapted to model the computation offloading process in a multi-mobile user, multi-edge server environment. This approach can help identify bottlenecks and optimize resource utilization by analyzing the arrival rates, service rates, and file sizes of computation tasks. However, the model would need to be extended to account for the dynamic and adaptive nature of the target problem, incorporating factors such as user mobility, edge server heterogeneity, and finite resources.
\\
\midrule
H\newline
{\small ranks 1 / 10}
&
\textit{What are the diverse roles of pericytes in disease pathobiology, and how can understanding these roles inform future therapeutic strategies?}
\par
Inspiration: \emph{A molecular atlas of cell types and zonation in the brain vasculature}
&
A detailed molecular characterization of pericytes using single-cell transcriptomics can reveal their diverse roles in disease pathobiology, including their involvement in fibrosis and immune modulation, and inform therapeutic strategies by identifying specific molecular profiles and functional roles that can be targeted for treatment.
&
One transferable mechanism principle from the inspiration paper is the use of single-cell transcriptomics to define the molecular characteristics of pericytes and their roles in different biological contexts. This approach can be adapted to study pericytes in various diseases and tissues, providing insights into their diverse roles and potential therapeutic targets. By applying this principle, researchers can gain a deeper understanding of pericyte biology and develop targeted therapeutic strategies for a range of disorders.
\\
\midrule
I\newline
{\small ranks 1 / 8}
&
\textit{How can the accuracy and reliability of physics-informed neural networks (PINNs) be improved for simulating problems involving strong nonlinear discontinuities, such as shock waves in hyperbolic equations?}
\par
Inspiration: \emph{Understanding and mitigating gradient flow pathologies in physics-informed neural networks}
&
A learning rate annealing algorithm that utilizes gradient statistics during model training can be employed to balance the interplay between different terms in the composite loss function of PINNs, thereby mitigating the effects of numerical stiffness and improving the stability and accuracy of PINNs in capturing strong nonlinear discontinuities in hyperbolic equations.
&
A potential mechanism for improving the accuracy and reliability of PINNs for simulating strong nonlinear discontinuities is to adapt the learning rate annealing algorithm to balance the interplay between different terms in composite loss functions. This could help mitigate the numerical stiffness issue and improve the predictive accuracy of PINNs for the target problem. However, a more robust approach may be needed to specifically address the challenges of simulating strong nonlinear discontinuities, such as shock waves in hyperbolic equations.
\\
\bottomrule
\end{tabular}
\caption{Additional TCA-SIR vs.\ Prompt-TCA generations (Examples~G--I; ranks: TCA-SIR / Prompt-TCA).
These cases show target-conditioned wording but often shared mechanism family; they were previously used as an earlier main-paper qualitative set.}
\label{tab:qual-supp-former-abc}
\end{table*}

%% file: tables/qualitative_supp_extra.tex
\begin{table*}[t]
\centering
\small
\setlength{\tabcolsep}{1mm}
\begin{tabular}{@{}p{0.16\textwidth}>{\raggedright\arraybackslash\hspace{0pt}}p{0.26\textwidth}>{\raggedright\arraybackslash\hspace{0pt}}p{0.26\textwidth}>{\raggedright\arraybackslash\hspace{0pt}}p{0.26\textwidth}@{}}
\toprule
& \textbf{TCA-SIR} & \textbf{Prompt-TCA} & \textbf{SourceAbs} \\
\midrule
D\newline
{\small ranks 2 / 11 / 3}
\par
{\small LAST + E-Walker assembly}
\par
{\small Insp.: space-robotics OOS}
&
A transferable principle for LAST assembly by an E-Walker is to use advanced microgravity control (nonholonomic / adaptive / robust) that limits base disturbances and stabilizes the manipulator after contact.
&
Develop advanced motion-planning and manipulation controllers for microgravity; scoped mainly to individual E-Walker components rather than full large-scale assembly coordination.
&
Integrate kinematics, dynamics, and control so robotic systems can handle uncertain dynamics, a holistic but less target-specific principle.
\\
\midrule
E\newline
{\small ranks 3 / 9 / 1}
\par
{\small SLC7A11 / disulfidptosis}
\par
{\small Insp.: disulfide switches}
&
Regulate disulfide bonds in SLC7A11 (including those tied to disulfidptosis) as cleavable switches for cancer therapeutics that modulate cell-death pathways.
&
Treat disulfide cleavage as a protein-function switch and ask how that lens applies to SLC7A11-regulated cystine transport / glutathione synthesis in disulfidptosis.
&
Reinterpret stable structural elements as dynamic regulatory switches that fine-tune protein activity under changing conditions.
\\
\midrule
F\newline
{\small ranks 1 / 9 / 2}
\par
{\small flame-retardant light SiRF}
\par
{\small Insp.: MXene retardants}
&
Adapt MXene-style nanomaterials to boost SiRF flame retardancy and thermal stability without heavy filler loadings that ruin lightness and flexibility.
&
Incorporate high-surface-area nanofillers (e.g., Ti$_3$C$_2$T$_x$) to disperse flame retardants in SiRF; notes remaining stability / aggregation questions.
&
Add functional nanomaterials to composites for multifunctionality, with dispersion and compatibility as the main engineering bottlenecks.
\\
\bottomrule
\end{tabular}
\caption{Additional model-generated abstractions (Examples~D--F; ranks: TCA-SIR / Prompt-TCA / SourceAbs).
D: cross-task space-robotics transfer (on-orbit servicing review $\rightarrow$ E-Walker telescope assembly).
E: general-to-specific disulfide mechanism (protein disulfide switches $\rightarrow$ SLC7A11/disulfidptosis); SourceAbs can rank first while remaining domain-washed.
F: materials transfer (MXene flame-retardant polymers $\rightarrow$ flame-retardant silicone rubber foam).}
\label{tab:qual-supp-extra}
\end{table*}

%% file: tables/qualitative_supp_paa.tex
\begin{table*}[t]
\centering
\small
\setlength{\tabcolsep}{1mm}
\begin{tabular}{p{0.32\textwidth}p{0.55\textwidth}}
\toprule
\textbf{Target focus} & \textbf{TCA-SIR abstraction (same inspiration)} \\
\midrule
Optimize a specific PAA@TiO$_2$ nanocomposite for crystal-violet removal
& Integrate experiments with DFT, MD, and MC to characterize adsorption mechanisms and thermodynamics, enabling optimization of the PAA@TiO$_2$ adsorbent.
\par
{\small Chemistry target; TCA-SIR rank~1}
\\
\midrule
Design an effective, cost-efficient adsorbent (molecular-level understanding secondary)
& Integrate nanoparticles with a polymer matrix to improve stability/dispersion and adsorption capacity for a sustainable water-treatment material.
\par
{\small Materials target; TCA-SIR rank~1}
\\
\bottomrule
\end{tabular}
\caption{Same inspiration, different targets (controlled target conditioning).
Shared source: \emph{Investigation of the anionic polyacrylamide as a potential adsorbent of crystal violet dye\ldots{} (DFT, MC, MD)}.
TCA-SIR emphasizes experimental--computational characterization when the target asks how to optimize PAA@TiO$_2$, but polymer--nanoparticle composite design when the target asks how to develop a broadly effective adsorbent.
The targets are related formulations of a similar problem, so this is preliminary evidence of facet selection rather than transfer across unrelated goals.}
\label{tab:qual-supp-paa}
\end{table*}

%% file: tables/qualitative_supp_genlevel.tex
\begin{table*}[t]
\centering
\small
\setlength{\tabcolsep}{1mm}
\begin{tabular}{@{}p{0.03\textwidth}p{0.45\textwidth}p{0.45\textwidth}@{}}
\toprule
& \textbf{Target L1--L4} & \textbf{GT inspiration L1--L4} \\
\midrule
G &
L1: Develop an optimization algorithm for computation offloading in a multi-mobile user, multi-edge server environment with heterogeneous and finite resources to minimize user time and energy consumption while maximizing server utilization.
\par L2: Design a multi-objective optimization framework for task allocation in distributed systems with heterogeneous resources and finite capacities to balance competing performance metrics.
\par L3: Create a resource allocation strategy for complex systems with multiple users and limited resources to optimize performance and efficiency while minimizing waste.
\par L4: Find an optimal allocation of limited resources among competing demands to maximize overall system performance and minimize waste.
&
Inspiration: \emph{A queuing theory model for cloud computing}
\par L1: Improving QoS in cloud computing environments using queuing theory.
\par L2: Analyzing and optimizing performance in distributed computing systems.
\par L3: Identifying and addressing bottlenecks in complex systems to enhance efficiency.
\par L4: Optimizing system performance by understanding and managing resource utilization and capacity.
\par {\small Best cosine pair: target L3 $\leftrightarrow$ source L4 (group rank~1; shortlist rank~2).}
 \\
\midrule
H &
L1: What are the diverse roles of pericytes in disease pathobiology?
\par L2: How do specialized cells contribute to vascular stability and function in various biological contexts?
\par L3: What are the mechanisms by which cells with diverse functions interact with their environment to influence disease progression?
\par L4: How do complex systems integrate diverse cell types to maintain homeostasis and respond to pathological conditions?
&
Inspiration: \emph{A molecular atlas of cell types and zonation in the brain vasculature}
\par L1: Understanding the molecular characteristics of brain vascular cells in the adult mouse brain.
\par L2: Characterizing the molecular properties of vascular cells in various organs and tissues to understand their roles in health and disease.
\par L3: Applying single-cell transcriptomics to identify and categorize cell types in complex biological systems.
\par L4: The importance of cellular heterogeneity in understanding complex biological systems and its implications for disease.
\par {\small Best cosine pair: target L2 $\leftrightarrow$ source L2 (group rank~1; shortlist rank~2).}
 \\
\midrule
I &
L1: How can the accuracy and reliability of PINNs be improved for simulating shock waves in hyperbolic equations?
\par L2: How can numerical methods be enhanced to accurately capture strong nonlinear discontinuities in complex systems governed by PDEs?
\par L3: How can machine learning models be adapted to effectively handle abrupt changes in physical quantities in various complex systems?
\par L4: How can models be improved to handle abrupt changes in complex systems with high uncertainty and nonlinearity?
&
Inspiration: \emph{Understanding and mitigating gradient flow pathologies in PINNs}
\par L1: Physics-informed neural networks in computational physics.
\par L2: Training constrained neural networks in scientific machine learning.
\par L3: Mitigating numerical instability in neural networks with composite loss functions.
\par L4: Balancing competing objectives in complex optimization problems.
\par {\small Best cosine pair: target L1 $\leftrightarrow$ source L3 (group rank~9; not shortlisted).}
 \\
\bottomrule
\end{tabular}
\caption{Gen-level retrieval intermediate abstractions on Examples~G--I (\texttt{max} variant).
Higher levels (esp.\ L3--L4) wash mechanism detail into broad ``complex systems'' / ``resource allocation'' language, so cosine matching can reward vague topical overlap (G,\,H) or miss the actionable mechanism transfer entirely (I).}
\label{tab:qual-supp-genlevel}
\end{table*}

%% file: tables/qualitative_supp_idea_catalyst.tex
\begin{table*}[t]
\centering
\small
\setlength{\tabcolsep}{1mm}
\begin{tabular}{@{}p{0.03\textwidth}p{0.30\textwidth}p{0.30\textwidth}p{0.30\textwidth}@{}}
\toprule
& \textbf{Target core problem} & \textbf{GT inspiration core problem} & \textbf{Illustrative distractor core} \\
\midrule
G &
How to efficiently allocate limited resources among competing tasks to minimize time and energy consumption while maximizing utilization.
&
Inspiration: \emph{A queuing theory model for cloud computing}
\par
Developing a framework to predict and optimize the performance of complex systems under varying loads and constraints.
&
Paper: \emph{Developing and validating trust measures for e-commerce}
\par
Developing a unified framework to measure complex, multifaceted concepts across different contexts.
\\
\midrule
H &
Understanding the dynamic behavior of specialized cells in complex biological systems to identify their diverse roles and potential therapeutic targets.
&
Inspiration: \emph{A molecular atlas of cell types and zonation in the brain vasculature}
\par
Identifying and categorizing the distinct characteristics of individual cell types within a complex system to understand their roles and relationships.
&
Paper: \emph{Gaia Data Release 2}
\par
Integrating and processing large amounts of diverse data to provide a comprehensive and accurate representation of a complex system.
\\
\midrule
I &
How to effectively handle abrupt changes in a system's behavior while maintaining stability and preserving underlying physical properties.
&
Inspiration: \emph{Understanding and mitigating gradient flow pathologies in PINNs}
\par
How to balance competing objectives in complex optimization problems to prevent numerical instability and improve overall performance.
&
Paper: \emph{SC3: consensus clustering of single-cell RNA-seq data}
\par
Identifying a robust and accurate method to group similar items into clusters when the data is noisy and multiple solutions are possible.
\\
\bottomrule
\end{tabular}
\caption{Idea-Catalyst intermediate core problems on Examples~G--I.
Target and source statements are intentionally domain-agnostic one-liners; as a result, unrelated papers can share nearly interchangeable ``complex system'' cores (G,\,H), and even when the GT inspiration is shortlisted the extracted problem rarely retains the transferable \emph{mechanism} that TCA-SIR surfaces.}
\label{tab:qual-supp-idea-catalyst}
\end{table*}

%% file: appendix/implementation_details.tex
\section{Implementation Notes}
\label{sec:supp-implementation}

\subsection{Ranking metrics}
HitRate follows ResearchBench~\citep{liu2026researchbench}: for each target, the fraction of gold inspirations among the top $k$ retrieved candidates, with $k{=}3$ (top~4\% of a 75-pool) or $k{=}15$ (top~20\%), then averaged over targets.
MRR is $1/r$ for the rank $r$ of the first gold inspiration ($0$ if none is retrieved), averaged over targets.
For NDCG@3, each of the top three positions contributes a binary gain of $1$ if it matches a still-unmatched gold inspiration, discounted by $\log_2(i{+}1)$ at position $i$, then normalized by the ideal top-3 ranking; we report the mean over targets together with a bootstrap 95\% CI half-width.

\subsection{Hardware and software}
All reported training and HitRate evaluations were run on a dual-socket Linux server with 512\,GB RAM and two NVIDIA RTX PRO~6000 GPUs (96\,GB each), under Ubuntu Server~24.04.
Each fine-tuning or evaluation job used a single GPU.
The reported TCA-SIR LoRA fine-tune (Llama-3.1-8B-Instruct, 3~epochs, learning rate $1{\times}10^{-5}$) takes roughly 8~hours wall-clock on one GPU; ablation and sensitivity runs are in the same regime.
Software stack: Python~3.11, PyTorch with CUDA, Hugging Face Transformers / PEFT / TRL, and the frozen YAML configs shipped with the code-and-data package.

\subsection{Artifacts and configs}
Frozen ResearchBench holdouts used in this paper live under the submission data directory:
\textbf{test240} (headline HitRate), \textbf{val120} (training-time validation / ablation screening),
and \textbf{pilot12} (one target per domain from val120, seed~42, for sensitivity sweeps).
Training and HitRate evaluation configs are the frozen YAML copies under the submission configs
directory, with per-experiment notes under the submission experiments directory.
Paper display names (TCA-SIR / Prompt-TCA / SourceAbs) are defined in the analysis naming module
and \texttt{NAMING.md} in the paper folder.
Tables and figures in this appendix can be regenerated from the repository root with the
submission \texttt{run\_paper\_assets.sh} script.

%% file: appendix/teacher_supervision_quality.tex
\section{Teacher Supervision Quality}
\label{sec:supp-teacher-supervision}

We inspected the accepted frozen supervision rows used for TCA-SIR training (\(n=3967\), from the same v4c text-anchor SFT set used in the main experiments) to check whether the 5-level grader collapses to an almost binary \(\{0.1,0.9\}\) assignment.
Figures~\ref{fig:supp-score-dist}--\ref{fig:supp-score-summary} show that it does not.
Positive pairs concentrate at \(0.5\) and \(0.7\), negatives concentrate at \(0.1\), and the two classes still overlap at intermediate levels rather than separating into a degenerate two-point code.
This is consistent with the grader adding transfer nuance beyond the original positive/negative pairing.

\newcommand{\scorehistcustom}[6]{
\begin{tikzpicture}[x=0.8cm,y=0.055cm]
  \node[font=\small] at (3.0, 93) {#1};
  \draw[black!60] (0,0) rectangle (6.1,82);
  \draw[->,black!70] (0.6,10) -- (5.7,10);
  \draw[->,black!70] (0.6,10) -- (0.6,76);
  \foreach \y in {0,20,40,60} {
    \draw[black!70] (0.52,\y+10) -- (0.68,\y+10);
    \node[font=\small, anchor=east] at (0.45,\y+10) {\y};
  }
  \node[font=\small] at (3.2,2.5) {Score level};
  \node[font=\small, rotate=90] at (0.08,43) {Percent};
  \foreach \x/\lab/\val in {1/0.1/#2,2/0.3/#3,3/0.5/#4,4/0.7/#5,5/0.9/#6} {
    \fill[blue!65] (\x+0.35,10) rectangle (\x+0.95,\val+10);
    \node[font=\small] at (\x+0.65,6.2) {\lab};
    \node[font=\small, anchor=south] at (\x+0.65,\val+11.2) {\val};
  }
\end{tikzpicture}
}

\begin{figure*}[t]
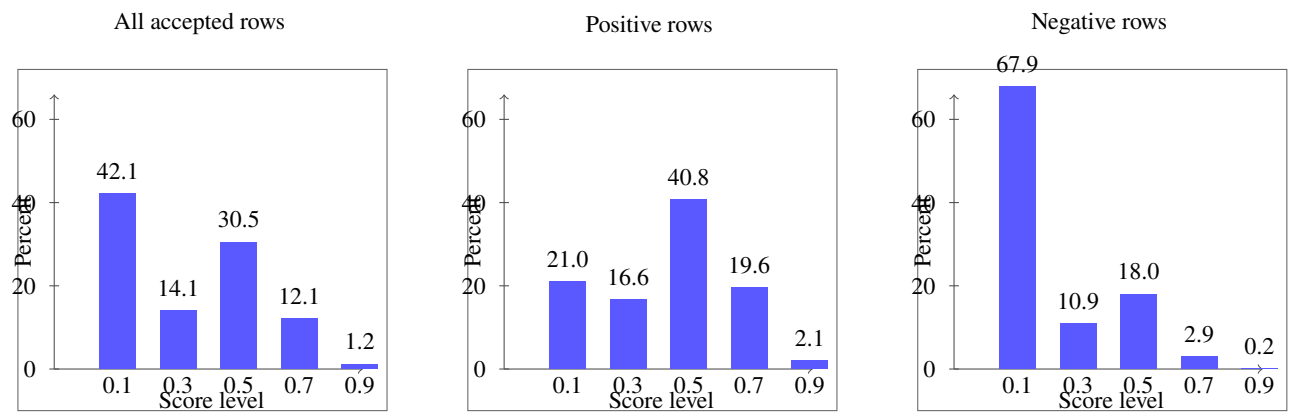

\centering
\scorehistcustom{All accepted rows}{42.1}{14.1}{30.5}{12.1}{1.2}\hspace{0.6cm}
\scorehistcustom{Positive rows}{21.0}{16.6}{40.8}{19.6}{2.1}\hspace{0.6cm}
\scorehistcustom{Negative rows}{67.9}{10.9}{18.0}{2.9}{0.2}
\caption{Five-level supervision distributions over accepted training rows. Left: overall distribution. Middle/right: per-class score histograms normalized within positive and negative rows, respectively. The mass is not collapsed to only \(0.1\) and \(0.9\): positives peak at \(0.5\) and \(0.7\), while negatives peak at \(0.1\) but still occupy \(0.3\) and \(0.5\).}
\label{fig:supp-score-dist}
\end{figure*}

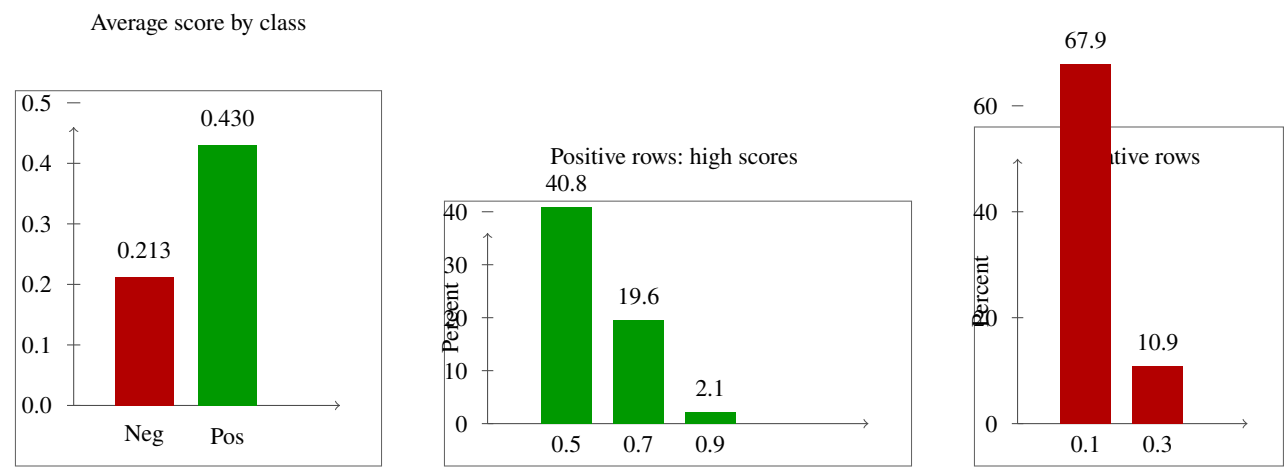
\begin{figure*}[t]
\centering
\begin{tikzpicture}[x=1.1cm,y=8cm]
  \node[font=\small] at (2.2, 0.73) {Average score by class};
  \draw[black!60] (0,0) rectangle (4.4,0.62);
  \draw[->,black!70] (0.7,0.10) -- (3.9,0.10);
  \draw[->,black!70] (0.7,0.10) -- (0.7,0.56);
  \foreach \y/\lab in {0/0.0,0.1/0.1,0.2/0.2,0.3/0.3,0.4/0.4,0.5/0.5} {
    \draw[black!70] (0.62,\y+0.10) -- (0.78,\y+0.10);
    \node[font=\small, anchor=east] at (0.55,\y+0.10) {\lab};
  }
  \foreach \x/\lab/\val/\col in {1/Neg/0.213/red!70!black,2/Pos/0.430/green!60!black} {
    \fill[\col] (\x+0.2,0.10) rectangle (\x+0.9,\val+0.10);
    \node[font=\small] at (\x+0.55,0.05) {\lab};
    \node[font=\small, anchor=south] at (\x+0.55,\val+0.115) {\val};
  }
\end{tikzpicture}
\hspace{0.5cm}
\begin{tikzpicture}[x=0.95cm,y=0.07cm]
  \node[font=\small] at (3.2, 58) {Positive rows: high scores};
  \draw[black!60] (0,0) rectangle (6.5,50);
  \draw[->,black!70] (0.6,8) -- (5.9,8);
  \draw[->,black!70] (0.6,8) -- (0.6,44);
  \foreach \y in {0,10,20,30,40} {
    \draw[black!70] (0.52,\y+8) -- (0.68,\y+8);
    \node[font=\small, anchor=east] at (0.45,\y+8) {\y};
  }
  \node[font=\small, rotate=90] at (0.08,28) {Percent};
  \foreach \x/\lab/\val in {1/0.5/40.8,2/0.7/19.6,3/0.9/2.1} {
    \fill[green!60!black] (\x+0.35,8) rectangle (\x+1.05,\val+8);
    \node[font=\small] at (\x+0.7,4.2) {\lab};
    \node[font=\small, anchor=south] at (\x+0.7,\val+9.2) {\val};
  }
\end{tikzpicture}
\hspace{0.5cm}
\begin{tikzpicture}[x=0.95cm,y=0.07cm]
  \node[font=\small] at (2.2, 58) {Negative rows};
  \draw[black!60] (0,0) rectangle (4.3,64);
  \draw[->,black!70] (0.6,8) -- (3.8,8);
  \draw[->,black!70] (0.6,8) -- (0.6,58);
  \foreach \y/\lab in {0/0,20/20,40/40,60/60} {
    \draw[black!70] (0.52,\y+8) -- (0.68,\y+8);
    \node[font=\small, anchor=east] at (0.45,\y+8) {\lab};
  }
  \node[font=\small, rotate=90] at (0.08,33) {Percent};
  \foreach \x/\lab/\val in {1/0.1/67.9,2/0.3/10.9} {
    \fill[red!70!black] (\x+0.2,8) rectangle (\x+0.9,\val+8);
    \node[font=\small] at (\x+0.55,4.2) {\lab};
    \node[font=\small, anchor=south] at (\x+0.55,\val+9.2) {\val};
  }
\end{tikzpicture}
\caption{Summary views of teacher-supervision quality. Left: average graded score by class (positive \(0.430\), negative \(0.213\)). Middle panels: within positive rows, \(40.8\%\) receive \(0.5\), \(19.6\%\) receive \(0.7\), and \(2.1\%\) receive \(0.9\). Right: within negative rows, \(67.9\%\) receive \(0.1\) while \(10.9\%\) receive \(0.3\). Together with Figure~\ref{fig:supp-score-dist}, this overlap supports a nuanced grader rather than a near-binary collapse.}
\label{fig:supp-score-summary}
\end{figure*}

%% file: appendix/prompts.tex
\section{Training-Data Construction Prompts}
\label{sec:supp-prompts}

We use Llama-3.1-8B-Instruct for both the label-aware R/A teacher and the transferability grader.
Placeholders of the form \texttt{\{problem\}} are filled from each ResearchBench pair at runtime.
The templates below are the production prompts for TCA-SIR supervision (v2 teacher templates and the 5-level grader).
Prompts are shown in a reduced monospace font and hard-wrapped to the column width.

\subsection{Positive teacher (label-aware R/A)}
{\small\begin{verbatim}
Given a target research problem and
an inspiration paper, produce a
transfer decision with reasoning and
abstraction.

Target problem:
{problem}

Background:
{background}

Inspiration title:
{title}

Inspiration abstract:
{abstract}

Set transfer_decision to
"transferable".
In reasoning (1-3 sentences), explain
which specific aspect of the
inspiration is useful for the target
problem.
In abstraction (one paragraph), state
a transferable principle adapted to
this target. Ground it in the
inspiration content.

Do NOT use markdown, bullet lists, or
generic filler. Do NOT copy refusal
boilerplate.

Respond with ONLY valid JSON (no
markdown fences):
{"transfer_decision": "transferable",
 "reasoning": "<1-3 sentences>",
 "abstraction": "<single paragraph>"}
\end{verbatim}}

\subsection{Negative teacher (scope-limited R/A)}
{\small\begin{verbatim}
Given a target research problem and
an inspiration paper, produce a
transfer decision with scope-limited
reasoning and abstraction.

Target problem:
{problem}

Background:
{background}

Inspiration title:
{title}

Inspiration abstract:
{abstract}

Set transfer_decision to
"not_transferable".
In reasoning (1-3 sentences), follow
this pattern:
"The inspiration appears related
because X, but it is limited to Y,
while the target requires Z.
Therefore, the transfer is
unreliable."
In abstraction (one paragraph), follow
this pattern:
"Scope-limited abstraction: the
source applies to Y and does not
provide a reliable transferable
principle for Z."
Name concrete X (superficial
overlap), Y (source scope), and Z
(target requirement).

Do NOT invent a positive bridge.
Do NOT use markdown or bullet lists.
Do NOT use generic refusal
boilerplate.

Respond with ONLY valid JSON (no
markdown fences):
{"transfer_decision":
 "not_transferable",
 "reasoning": "<1-3 sentences>",
 "abstraction": "<single paragraph>"}
\end{verbatim}}

\subsection{Five-level transferability grader}
{\small\begin{verbatim}
You are grading scientific inspiration
transferability.

Your task is to assign a graded
transferability score for how useful
the candidate inspiration paper is for
solving the target research problem.

You must judge the candidate based on
transferable mechanism, not surface
similarity.

Important distinction:
* A paper can be topically related but
  still weak if it does not provide a
  useful mechanism for the target.
* A paper can be from a different
  domain but strong if it provides a
  mechanism, method, or design
  principle that can be adapted to
  the target.
* A broad analogy is not enough for a
  high score.
* Do not overclaim transferability.
* Penalize candidates that are only
  generally related, only share
  keywords, or only provide background
  knowledge.

Scoring scale:
0.9 = Strong / central transfer. The
  candidate provides a mechanism,
  method, or design principle that
  directly addresses the target's
  main bottleneck.
0.7 = Useful partial transfer. Useful
  for part of the target, but does
  not fully address the central
  problem.
0.5 = Plausible broad analogy.
  Reasonable abstract connection, but
  the mechanism is generic, indirect,
  or not clearly central.
0.3 = Weak / scope-limited relation.
  Peripheral or requires major
  adaptation.
0.1 = Unrelated or misleading. No
  useful transferable mechanism.

Target research problem:
{target_problem}

Target background:
{target_background}

Candidate inspiration paper:
Title: {candidate_title}
Abstract: {candidate_abstract}

Original dataset metadata, for
reference only. Do not copy this
label. Re-grade independently:
{original_metadata}

Now grade the candidate.

Return valid JSON only. Do not
include markdown.

JSON schema:
{"graded_transfer_score": 0.1,
 "transfer_level":
   "unrelated | weak |
    broad_plausible | partial |
    strong",
 "label_rationale": "...",
 "central_transfer_mechanism": "...",
 "why_not_higher": "...",
 "overclaim_risk":
   "low | medium | high"}
\end{verbatim}}

%% file: appendix/training_examples.tex
\section{Training Supervision Examples}
\label{sec:supp-training-examples}

A natural concern is that label-aware R/A teaching could imprint a small set of polarity or grade-specific surface templates, and that a score head reading the final-token hidden state after generation could then recover the graded label from those templates rather than from transfer content.
Inspection of accepted training rows does not support a unique phrase-to-score mapping: within each graded score, Reasoning/Abstraction vary in framing (mechanism adaptation, methodological template, scope mismatch, domain-specific mechanism claims, etc.).
Below we show \emph{two} accepted training rows for each score in $\{0.1,0.3,0.5,0.7,0.9\}$ (verbatim teacher R/A; problem text truncated to the lead question).
Comparing the pair at a fixed score makes the within-label stylistic diversity visible.

\subsection{Graded $0.1$}
\paragraph{Example A (Astronomy; gold negative).}
\textbf{Target.}
How can artificial intelligence (AI) be effectively integrated into gravitational wave (GW) data analysis to enhance the efficiency, accuracy, and scope of scientific discoveries?

\textbf{Inspiration.}
\emph{Advanced Virgo: a second-generation interferometric gravitational wave detector.}

\textbf{Teacher Reasoning.}
The inspiration paper focuses on the design and construction of a gravitational wave detector, Advanced Virgo, which is a hardware upgrade rather than a software solution.
This paper does not address the integration of AI into gravitational wave data analysis, which is the target problem.
The paper's scope is limited to the detector's hardware and its expected performance, rather than software applications.

\textbf{Teacher Abstraction.}
The Advanced Virgo detector is a second-generation interferometric gravitational wave detector being constructed to increase the detection rate by three orders of magnitude, with a focus on hardware upgrades and detector performance, assuming a specific scale of detector size and sensitivity, and a method of interferometric detection.

\paragraph{Example B (Astronomy; gold positive).}
\textbf{Target.}
How can the significant imbalance between non-pulsar and pulsar samples in pulsar candidate datasets be effectively addressed to improve the accuracy of pulsar candidate identification using deep learning models?

\textbf{Inspiration.}
\emph{Neural discrete representation learning.}

\textbf{Teacher Reasoning.}
The inspiration's use of vector quantisation (VQ) to learn discrete representations can be applied to address the imbalance between non-pulsar and pulsar samples in pulsar candidate datasets by generating synthetic pulsar samples that can help alleviate the challenges of imbalanced datasets.

\textbf{Teacher Abstraction.}
A VQ-based generative model can be adapted to learn discrete representations of pulsar signals, enabling the synthesis of realistic pulsar samples that can be used to augment the training dataset and improve the performance of machine learning classifiers in pulsar candidate identification, thereby addressing the significant imbalance between non-pulsar and pulsar samples.

\subsection{Graded $0.3$}
\paragraph{Example A (Biology; gold positive).}
\textbf{Target.}
How can recent advances and clinical trials in pancreatic ductal adenocarcinoma (PDAC) improve patient outcomes through novel therapeutic strategies, biomarkers, and clinical trial designs?

\textbf{Inspiration.}
\emph{Master protocols: new directions in drug discovery.}

\textbf{Teacher Reasoning.}
The inspiration's concept of master protocols, particularly umbrella trials, can be applied to PDAC to evaluate multiple therapeutic options simultaneously, streamlining the drug development process and enabling personalized treatment approaches based on tumor biology.

\textbf{Teacher Abstraction.}
A master protocol design, specifically an umbrella trial, can be adapted for PDAC to evaluate multiple therapeutic combinations and biomarkers in a single study, leveraging molecular profiling to stratify patients and optimize treatment outcomes, thereby enhancing the efficiency and effectiveness of clinical trials in this challenging disease.

\paragraph{Example B (Biology; gold positive).}
\textbf{Target.}
How can recent advances and clinical trials in pancreatic ductal adenocarcinoma (PDAC) improve patient outcomes through novel therapeutic strategies, biomarkers, and clinical trial designs?

\textbf{Inspiration.}
\emph{Maintenance olaparib for germline BRCA-mutated metastatic pancreatic cancer.}

\textbf{Teacher Reasoning.}
The inspiration's use of PARP inhibitors as a targeted therapy for germline BRCA-mutated metastatic pancreatic cancer provides a useful approach for addressing the molecular heterogeneity and tumor vulnerabilities in PDAC, particularly in patients with specific genetic mutations.

\textbf{Teacher Abstraction.}
Targeted therapies that exploit specific tumor vulnerabilities, such as PARP inhibitors for BRCA-mutated PDAC, can improve patient outcomes by enhancing treatment efficacy and reducing adverse effects, underscoring the importance of integrating molecular data into clinical decision-making to personalize treatment approaches and address the complex molecular landscape of PDAC.

\subsection{Graded $0.5$}
\paragraph{Example A (Astronomy; gold positive).}
\textbf{Target.}
How can we improve the calibration of Adaptive Optics (AO) systems in telescopes to effectively compensate for static optical aberrations without modifying the optical path, thereby enhancing image quality?

\textbf{Inspiration.}
\emph{Phase retrieval and diversity in adaptive optics.}

\textbf{Teacher Reasoning.}
The inspiration's concept of phase retrieval and diversity in adaptive optics can be applied to improve the calibration of AO systems by leveraging the modulus of the signal from a system's focal plane array to estimate phase aberrations, which can be used to derive control signals to align the optical system.

\textbf{Teacher Abstraction.}
A transferable principle for improving AO system calibration is to utilize phase retrieval algorithms that estimate phase aberrations from the modulus of the signal from a system's focal plane array, enabling the derivation of control signals to dynamically adjust the optical system and compensate for static optical aberrations without modifying the optical path.

\paragraph{Example B (Biology; gold positive).}
\textbf{Target.}
How can a comprehensive understanding of the cellular and molecular mechanisms involved in wound healing lead to the development of novel therapeutics that prevent scarring and promote the repair of chronic wounds?

\textbf{Inspiration.}
\emph{Live imaging of wound angiogenesis reveals macrophage orchestrated vessel sprouting and regression.}

\textbf{Teacher Reasoning.}
The inspiration's findings on the crucial role of macrophages in orchestrating vessel sprouting and regression during wound healing provide a valuable insight into the complex cellular interactions involved in wound repair, which can be leveraged to develop novel therapeutics that prevent scarring and promote healing.

\textbf{Teacher Abstraction.}
A therapeutic strategy that modulates macrophage activation state and function could potentially enhance wound healing by promoting angiogenesis and preventing excessive scarring, as macrophages play a pivotal role in driving vessel sprouting and regression, and their dysregulation can lead to impaired healing outcomes.

\subsection{Graded $0.7$}
\paragraph{Example A (Cell Biology; gold positive).}
\textbf{Target.}
How can we significantly increase the throughput of long-read RNA sequencing to enable more comprehensive identification and quantification of full-length RNA isoforms, particularly in single-cell applications?

\textbf{Inspiration.}
\emph{ConcatSeq: a method for increasing throughput of single molecule sequencing by concatenating short DNA fragments.}

\textbf{Teacher Reasoning.}
The inspiration's ConcatSeq method, which concatenates short DNA fragments to increase the throughput of single molecule sequencing, can be adapted to long-read RNA sequencing by concatenating short RNA fragments to generate longer templates for sequencing, thereby increasing the throughput of long-read RNA sequencing.

\textbf{Teacher Abstraction.}
A transferable principle, dubbed Concatenate-and-Sequence, involves generating longer RNA templates by concatenating short RNA fragments, leveraging the capabilities of long-read sequencing technologies to increase the throughput and accuracy of full-length RNA isoform identification, particularly in single-cell applications.

\paragraph{Example B (Biology; gold positive).}
\textbf{Target.}
How can bispecific antibodies (bsAbs) be effectively developed and utilized in cancer therapy to achieve novel mechanisms of action and therapeutic applications that surpass those of conventional IgG-based antibodies?

\textbf{Inspiration.}
\emph{Biology drives the discovery of bispecific antibodies as innovative therapeutics.}

\textbf{Teacher Reasoning.}
The inspiration paper highlights the potential of bispecific antibodies to enable novel mechanisms of action and therapeutic applications by binding two different targets or epitopes, and the classification system for bsAb formats based on antigen-binding domains and Fc domain presence provides a framework for selecting a bispecific format to meet biology needs.

\textbf{Teacher Abstraction.}
A systematic approach to designing bispecific antibodies for cancer therapy involves categorizing formats based on antigen-binding domains and Fc domain presence to select a format that aligns with the specific biology of the target, thereby enabling novel mechanisms of action and therapeutic applications that surpass those of conventional IgG-based antibodies.

\subsection{Graded $0.9$}
\paragraph{Example A (Chemistry; gold positive).}
\textbf{Target.}
How can the structural stability and reversible capacity of P2-type layered transition metal oxides, used as cathode materials in sodium-ion batteries, be improved to achieve high energy density and long-cycle life?

\textbf{Inspiration.}
\emph{Identifying the critical role of Li substitution in P2-Na$_x$[Li$_y$Ni$_z$Mn$_{1-y-z}$]O$_2$ ($0 < x, y, z < 1$) intercalation cathode materials.}

\textbf{Teacher Reasoning.}
The inspiration paper highlights the beneficial effects of Li substitution in P2-type layered transition metal oxides, specifically the stabilization of the P2 structure and delay of phase transformation at high voltages, which is crucial for improving the structural stability and reversible capacity of these materials.

\textbf{Teacher Abstraction.}
Introducing Li substitution in P2-type layered transition metal oxides can stabilize the P2 structure and delay phase transformation at high voltages, allowing for more Na ions to reside in prismatic sites and maintaining the overall charge balance of the compound, thus enhancing the structural stability and reversible capacity of these materials.

\paragraph{Example B (Cell Biology; gold positive).}
\textbf{Target.}
How can we design advanced catalysts with optimized heterostructure interfaces to improve the performance of hydrogen electrocatalysis reactions, particularly in alkaline media?

\textbf{Inspiration.}
\emph{Interface engineering of heterostructured electrocatalysts towards efficient alkaline hydrogen electrocatalysis.}

\textbf{Teacher Reasoning.}
The inspiration paper highlights the importance of interface engineering in enhancing the performance of heterostructured electrocatalysts for hydrogen evolution and oxidation reactions in alkaline media, particularly through the promotion of reaction kinetics and the optimization of electronic structures at the interface.

\textbf{Teacher Abstraction.}
Designing advanced catalysts with optimized heterostructure interfaces for hydrogen electrocatalysis in alkaline media involves leveraging the synergistic effects of combining different materials at the interface to enhance electrocatalytic performance, while also optimizing the binding energies of key intermediates through electronic interaction, interfacial bonding, and lattice strain, ultimately leading to improved reaction kinetics and overall efficiency.